# DMC-ICE13: ambient and high pressure polymorphs of ice from Diffusion Monte Carlo and Density Functional Theory

Flaviano Della Pia,[1] Andrea Zen,[2,3] Dario Alfè,[2,3,4,5] and Angelos Michaelides[1]
[1)] *Yusuf Hamied Department of Chemistry, University of Cambridge, Cambridge CB2 1EW, United Kingdom*
[2)] *Dipartimento di Fisica Ettore Pancini, Università di Napoli Federico II, Monte S. Angelo, I-80126 Napoli, Italy*
[3)] *Department of Earth Sciences, University College London, London WC1E 6BT, United Kingdom*
[4)] *Thomas Young Centre, University College London, London WC1E 6BT, United Kingdom*
[5)] *London Centre for Nanotechnology, University College London, London WC1E 6BT, United Kingdom*

Ice is one of the most important and interesting molecular crystals exhibiting a rich and evolving phase diagram. Recent discoveries mean that there are now twenty distinct polymorphs; a structural diversity that arises from a delicate interplay of hydrogen bonding and van der Waals dispersion forces. This wealth of structures provides a stern test of electronic structure theories, with Density Functional Theory (DFT) often not able to accurately characterise the relative energies of the various ice polymorphs. Thanks to recent advances that enable the accurate and efficient treatment of molecular crystals with Diffusion Monte Carlo (DMC), we present here the DMC-ICE13 dataset; a dataset of lattice energies of 13 ice polymorphs. This dataset encompasses the full structural complexity found in the ambient and high-pressure molecular ice polymorphs and when experimental reference energies are available our DMC results deliver sub-chemical accuracy. Using this dataset we then perform an extensive benchmark of a broad range of DFT functionals. Of the functionals considered, we find revPBE-D3 and RSCAN to reproduce reference absolute lattice energies with the smallest error, whilst optB86b-vdW and SCAN+rVV10 have the best performance on the relative lattice energies. Our results suggest that a single functional achieving reliable performance for all phases is still missing, and that care is needed in the selection of the most appropriate functional for the desired application. The insights obtained here may also be relevant to liquid water and other hydrogen bonded and dispersion bonded molecular crystals.

## I. INTRODUCTION

Water and ice are ubiquitous in nature and relevant to an almost endless list of scientific problems in materials science, chemistry, physics, and biology. This broad interest has motivated many decades of research into the phase diagram of water and ice, with renewed interest in the last few years owing for example to the discovery of several new crystalline polymorphs[1–5], and the theoretical prediction of numerous further candidates, e.g. Ref. 6. The recent experimental discoveries have increased the already extreme complexity of the water phase diagram to 20 solid phases, the liquid state, and various amorphous phases[7–11].

An accurate description of the phase diagram has been a major challenge for computational approaches and excellent progress has been made both with classical potentials[12,13] and with Machine Learning (ML) models[14,15]. The ML work have been particularly impressive, testing several Density Functional Theory (DFT) functionals (SCAN, B3LYP+D, PBE0+D and revPBE0+D) as well as taking into account quantum nuclear effects (QNEs). However, differences between the computed and experimental phase boundaries still exist. Although thermal and QNEs are important, the key to the phase diagram is an accurate description of the relative energies of the various ice polymorphs. It is this issue that the current article focuses on.

The main parameters used to assess the stability of ice polymorphs are the absolute lattice energy (i.e. the crystal total energy relative to gas phase water molecules) and the relative lattice energy (i.e. the lattice energy of a polymorph relative to hexagonal ice). Since the first DFT study of Hamann[16] in 1997, a large number of DFT studies (considering a broad range of exchange-correlation functionals) have followed (see e.g. Refs. 17–23 and for a review see Ref. 24). Considerable insight has emerged from these studies, notably the realisation that non-local van der Waals (vdW) dispersion forces need to be accounted for an accurate description of the relative energies of the different polymorphs.

Generally the above studies have made reference to benchmark lattice energies derived from experiment in the 1984 work of Whalley[25]. Consequently benchmark lattice energies are available only for polymorphs characterized before then. In the absence of experiment, high-level electronic structure theories can in principle provide an alternative source of benchmark reference data. And, indeed, the energetics of several ice polymorphs have been examined with various explicitly correlated electronic structure theories[26–29]. In particular, Diffusion Monte Carlo (DMC) gives excellent results in the computation of water-ice lattice energies[18,26,30]. However, largely because of the computational cost of DMC, estimates are only available for four polymorphs (ice Ih, II, VIII, and XI).

In this study, thanks to recent developments[31,32] enabling accurate and efficient DMC simulations for molecular crystals[26], we perform an extensive DMC study of ice polymorphs. High accuracy reference values of the lattice energy for 13 ice polymorphs are provided; with the 13 polymorphs selected to provide a broad treatment of the main bonding topologies found in the ambient and high pressure molecular ice polymorphs. We subsequently use the DMC reference energies to conduct a benchmark on a broad range of DFT exchange-correlation (XC) functionals. This reveals that different functionals perform better in capturing the stability with respect to the gas or the solid phase, indicating that care should be taken in selecting the optimal functional for a particular study. From another perspective, our results suggest that a "universal" functional for water, i.e. a functional that gives a good agreement for all phases, is still missing.

The outline of the manuscript is as follows. In Sec. II we introduce the 13 ice polymorphs under consideration, then give technical details of both DFT and DMC simulations. In Sec. III we first present the DMC reference values, comparing with available experimental results. Subsequently, we present the outcome of the benchmark of numerous DFT-XC functionals, analysing the stability of the considered polymorphs with respect to the gas phase and hexagonal ice. Finally, we give some concluding remarks in Sec. IV.

## II. MATERIALS AND METHODS

### A. Dataset setup

In this work we compiled the `DMC-ICE13` set by considering 13 ice polymorphs, including hydrogen ordered and disordered phases that span a broad range of temperatures and pressures on the water phase diagram[11]. First, we considered the structures previously included in the DFT database `ICE10`[21] - ice Ih, II, III, VI, VII, VIII, IX, XIII, XIV, and XV. Then, we added the ordered counterpart of hexagonal ice, ice XI, leaving out the cubic ice Ic, expected to be isoenergetic with ice XI/Ih within the DMC statistical error[30]. Finally, we considered the meta-stable and self-interpenetrating structure of ice IV, and the recently discovered ice XVII[3]. Interestingly, this is an ultra-low-density porous state and has its stability domain in the negative pressure regions, but is meta-stable at ambient pressures and low temperatures. In this way, we are taking into account at least one polymorph for each hydrogen ordered-disordered couple (both for ice XI/Ih, VIII/VII, IX/III, and XV/VI). We leave out of the set ice XII and V, disordered counterparts of ice XIV and XIII, respectively. Finally, we do not include the least dense ice XVI, stable only at negative pressures, ice XIX, described as the ‘glassy’ counterpart of ice VI, and the high pressures symmetric ice X and ‘superionic water’ ice XVIII. In fact, these phases go beyond the intent of this study, which is focused on molecular crystals.

We note here that, in principle, several structures should be considered when computing the lattice energy of hydrogen-disordered polymorphs. Based on a DFT analysis reported in the SI, we estimate that the differences among the lattice energies of different hydrogen arrangements are of the order of the DMC error bars. Being indistinguishable at the DMC level, we consider only one hydrogen arrangement for each hydrogen-disordered phase.

The geometries of the considered structures are reported in the SI and shown in Fig. 1. Input DFT and DMC files are provided as Supplementary Material, to facilitate accessibility and reproducibility of our data.

### B. Geometry optimisation and DFT lattice energies

The physical quantity usually considered to establish the stability of a crystal is its absolute lattice energy, which is the energy per molecule gained upon assuming the crystal form with respect to the gas phase. It can be computed as

$$E_{\mathrm{latt}} = E_{\mathrm{crys}} - E_{\mathrm{gas}}, \tag{1}$$

where $E_{\mathrm{crys}}$ is the energy per molecule in the crystal phase, and $E_{\mathrm{gas}}$ is the energy of the isolated molecule. However, we are also interested in capturing the relative stability of the ice polymorphs, i.e. the stability with respect to a fixed crystalline phase instead of the gas state. This property is more relevant in, e.g. the computation of the water phase diagram. Therefore, we assess the relative stability of the crystalline phases by computing the relative lattice energy with respect to hexagonal ice Ih. For a general polymorph ‘$x$’, this is simply computed as

$$\Delta E^{x}_{\mathrm{latt}} = E^{x}_{\mathrm{latt}} - E^{\mathrm{Ih}}_{\mathrm{latt}}, \tag{2}$$

and is independent of the configuration of the monomer in the gas phase.

Initial structures of ice Ih, II, VIII, XIII, XIV, and XV were taken from Ref. 18, ice III, VI, VII, IX, and XVII were taken from Ref. 6, ice IV was taken from Ref. 34, and ice XI from Ref. 30. DFT calculations have been performed with the VASP program package[35–38]. The projector-augmented plane wave method (PAW) has been used with hard pseudo-potentials[39,40], with a dense FFT grid and a PAW energy cut-off of 1000 eV, necessary to achieve convergence in unconstrained geometry optimisation, as reported in Ref. 21. To be consistent in the benchmark of several exchange-correlation functionals with respect to the reference (DMC) value computed on a fixed geometry, we relaxed all the ice polymorph structures at a fixed DFT-XC level. These structures have been therefore used both in the DMC and DFT evaluations of the lattice energies. This is the standard procedure also adopted in previous benchmarks[18,19,21].

Since previous DMC calculations on ice Ih, II, and VIII, with PBE structures have shown results in excellent agreement with experimental values[18,26], we decided to optimise the geometry of all the considered polymorphs with the PBE functional; the potential error due to this approximation is estimated to be smaller than $\sim 1$ kJ/mol, as discussed in the SI. All geometries have been carefully optimised until all forces were less than $\sim 0.002$ eV/$\mathring{\mathrm{A}}^3$, sampling the Brillouin zone with a $3 \times 3 \times 3$ **k**-point grid centred on the $\Gamma$ point. The convergence of the structure relaxation for several polymorphs has been subsequently checked using a denser $5 \times 5 \times 5$ grid. A $3 \times 3 \times 3$ **k**-point grid has been used to perform GGAs, meta-GGAs, and vdW-inclusive single-point periodic calculations, yielding a 1 meV convergence threshold on the lattice energy, evaluated with respect to the $5 \times 5 \times 5$ grid. A denser $4 \times 4 \times 4$ **k**-point grid has been used for hybrid-XC calculations to achieve the same accuracy. The geometry used for the isolated molecule (gas phase) can also influence the value of $E_{\mathrm{latt}}$ (see Eq. 1). For this we used the most accurate geometry for the water monomer, derived by Partridge and Schwenke[41] using CCSD(T) and already used in previous analyses[18,26,42]. An additional DMC calculation has been performed for the computation of $E_{\mathrm{gas}}$ on the water monomer geometry optimised at the DFT-PBE level, leading to a variation in the absolute lattice energies smaller than 1 kJ/mol.

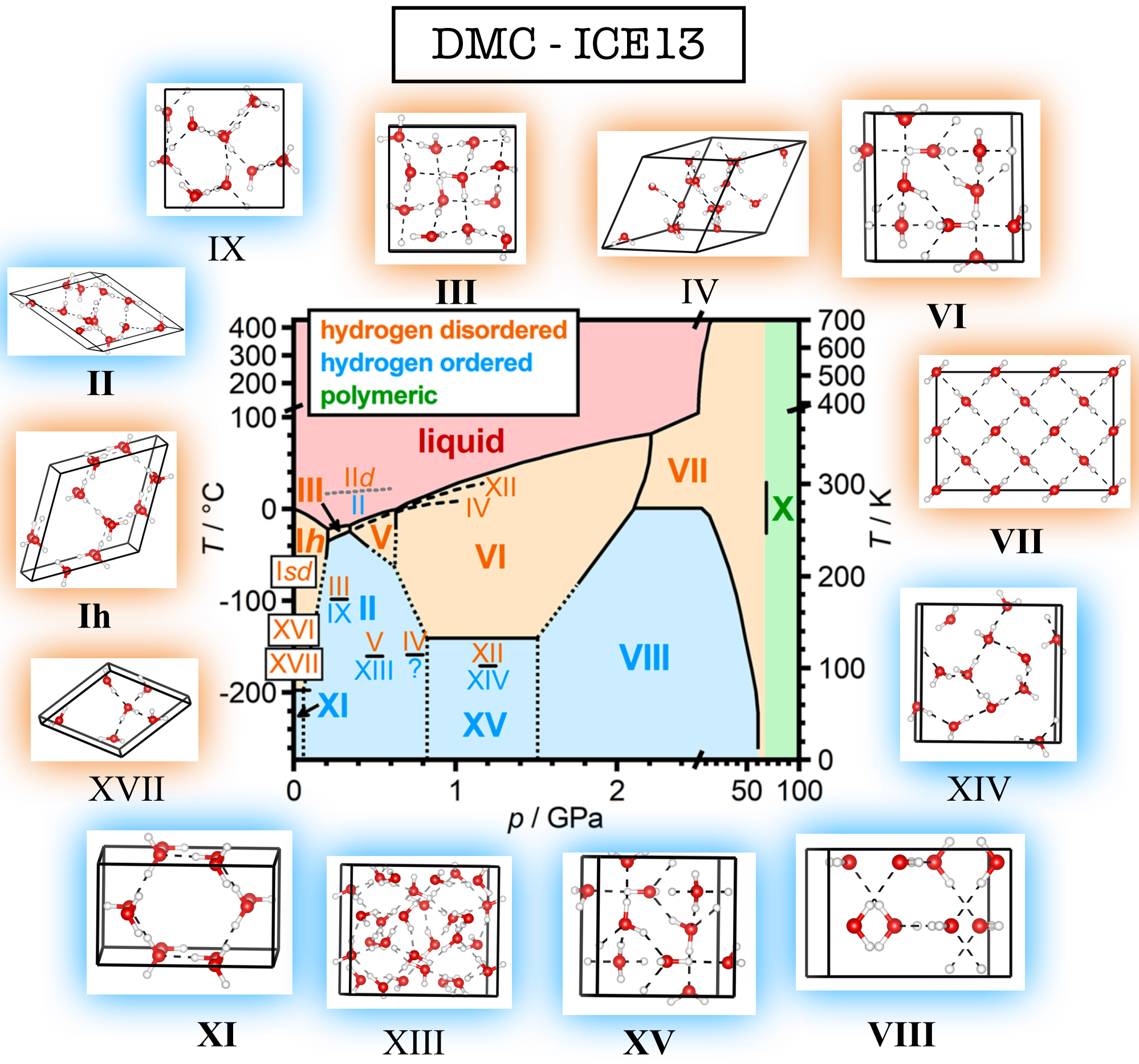

FIG. 1. Crystalline structures of the systems contained in the `DMC-ICE13` dataset. Each subplot reports the unit cell of the specified polymorph (plotted using VESTA[33]), with oxygen atoms in red, hydrogen atoms in white, and hydrogen-bonds as dashed black lines. The experimental phase diagram published in Ref. 11 is reported here to facilitate visualisation of the phase diagram regions covered by our dataset. Adopting the convention of Ref. 11, stable phases are indicated by large bold Roman numerals, whereas meta-stable states are indicated by a smaller font size. The phase diagram is reproduced from Ref. 11 with the permission of AIP Publishing.

There are now countless DFT functionals and we cannot test all or even most of them. Rather, here we focus on evaluating the performance of some of the most widely used families of functionals for water and ice. Specifically, we considered LDA[43], GGA (PBE[44], revPBE[45]), several dispersion-inclusive functionals (optB88-vdW[46], optB86-vdW[47], optPBE-vdW[46], vdW-DF[48], vdW-DF2[49], rev-vdW-DF2[50]), meta-GGA (SCAN[51], RSCAN[52], R2SCAN[53], SCAN+rVV10[54]), and hybrid methods (PBE0[55,56], revPBE0[57], B3LYP[58,59]). GGA and hybrid methods have also been applied with the D3 - with both zero and Becke-Johnson (BJ) damping, and with or without the three-body dispersion term of Axilrod-Teller-Muto (ATM) - and D4 London dispersion correction using the `dftd3/dftd4`[60–64] tools. PBE and PBE0 have also been evaluated with the Tkatchenko-Scheffler (TS) dispersion correction[65] and the Many-Body-Dispersion (MBD) method[66,67]. We also tested the Hartree-Fock (HF) approach.

### C. Diffusion Monte Carlo

Reference values for the lattice energies were computed with fixed-node DMC (FN-DMC), using the CASINO code[68]. We used Hartree-Fock pseudo-potentials[69,70] with the most recent determinant locality approximation (DLA)[32]. The trial wave-functions were of the Slater-Jastrow type with single Slater determinants, and the single-particle orbitals obtained from DFT local-density approximation (LDA) plane-wave calculations performed with PWscf[71,72] using an energy cut-off of 600 Ry and re-expanded in terms of B-splines[73]. The Jastrow factor included a two-body electron-electron (e-e) term, two-body electron-nucleus (e-n) terms, and three-body electron-electron-nucleus (e-e-n) terms. The variational parameters of the Jastrow have been optimised by minimising the variance in the simulated cell for each analysed polymorph. The size of the simulation cell imposes some constraints on the Jastrow variational freedom, in the form of cut-offs in the e-n, e-e and e-e-n terms. Following the workflow given in Ref. 26, tested on several molecular crystals including three ice polymorphs, the simulation cells have been generally defined in order to guarantee the radius of the sphere inscribed in the Wigner-Seitz cell to be bigger than 5 Å. The number of molecules in the simulated cell of each polymorph is reported in the SI.

The time step $\tau$ is a key issue affecting the accuracy of DMC calculations. In DMC, a propagation according to the imaginary time Schrödinger equation is performed to project out the exact ground state from a trial wave-function[74]. A time step $\tau$ must be chosen, but the projection is exact only in the continuous limit $\tau \to 0$. However, the ZSGMA[31] DMC algorithm gives better convergence with respect to $\tau$ than previously used methods, because the time-step bias per molecule is independent of the size of the simulated cell in a molecular crystals[26]. In this work we have verified the time step convergence for each analysed ice polymorph, as reported in the SI. We note that, in general, even in the limit of zero time step the DMC energy may be biased by the choice of the Jastrow factor, depending on how the non-local part of the pseudopotential is treated. This bias is eliminated if the DLA scheme is employed.

The computation of $E_{\rm crys}$ involves the use of periodic boundary conditions that can be subject to significant finite size errors (FSE). We took into account FSE using the Model Periodic Coulomb (MPC)[75–77] correction, and further correct for the (smaller) Independent Particle FSE (IPFSE) according to the procedure described in Ref. 26. An analysis of the FSE for each considered polymorph can be found in the SI.

Finally, the periodic DMC simulations have been performed using Twist Averaging Boundary Condition (TABC)[78]. This involves averaging the absolute DMC energies obtained using DFT-LDA single Slater determinants computed at different **k**-points in the Brillouin zone of the simulated cell. In particular, this means that the general DMC wave function of a periodic system is a complex function, except for a finite set of points in the Brillouin zone that make the wave function real, such as the $\Gamma$ point or the corner points. In this case, the FN-DMC is substituted by the non-equivalent fixed-phase DMC (FP-DMC). In principle, this means that also $E_{\rm gas}$ has to be computed with the FP-DMC, in order to be consistent in the estimation of $E_{\rm latt}$. However, it has been shown in Ref. 26 that the difference between the FN and FP estimates of $E_{\rm gas}$ for water is smaller than the statistical accuracy. For this reason, we used a real wave function in the DMC simulation of the gas phase.

## III. RESULTS AND DISCUSSION

The computation of the lattice energy is performed at zero temperature and pressure, and considering only the electronic contribution, i.e. neglecting quantum nuclear effects. Experimental estimates of the ice polymorphs lattice energies have been deducted from measures of the internal energy variation, according to an approximation described in the SI. These experimental values are affected by two types of error: an uncertainty coming from the actual measurement of the internal energy variation ($\sim 0.1$ kJ/mol), and an error due to the correction for the zero point energy effects, that we estimate to be of the order of $\sim 1-2$ kJ/mol, as generally found for molecular crystals in Ref. 26, and further analysed for ice polymorphs in the SI. This uncertainty on the experimental estimates of the lattice energies, as well as the lack of them for the recently discovered phases, are key reasons why results from a high accuracy electronic-structure method, such as DMC, are needed.

In Fig. 2 we report the DMC estimates, as well as the experimental values when available, of both the absolute (top panel) and relative (bottom panel) lattice energies for all the considered polymorphs. Exact values are also reported in Table I. First, we note that our DMC values for ice Ih and II agree within statistical error with the previous DMC results reported in Ref. 26. For ice VIII a small ($\sim 1.5$ kJ/mol) difference is found, which arises from the use, in the current study, of a bigger supercell and of the DLA in the pseudo-potentials (see Sec. II). Recently, high-accuracy Random Phase Approximation with exchange (RPAx)[28] lattice energies were computed for several crystalline systems, including ice II, VIII, IX and XI. We find our DMC estimates in good agreement with the reported values, with a maximum discrepancy of $\sim 1.2$ kJ/mol that can be mainly ascribed to the different functional (PBE+TS) used for the geometry optimisation.

Looking at the entire dataset, it is evident that the energy differences we want to catch in this work are minimal, in fact they are in most cases smaller than the 'chemical-accuracy' limit of 4 kJ/mol. In particular, ice XIV, XV, and XVII are degenerate within the DMC statistical error. Overall, all the lattice energies vary in a small range of 5 kJ/mol, defined by the lowest pressure phases ice Ih/XI and the highest pressure phase ice VII. Interestingly, we find that all the 'recent' polymorphs - where no experimental value is available - fall in this range. Overall, DMC is always in good agreement with the experimental values, with a maximum disagreement of 0.5 kJ/mol on the absolute lattice energies, and $\sim 1$ kJ/mol on the relative lattice energies. Thus, it ultimately defines the reference method we use to establish the performance of the DFT functionals.

The benchmark of several DFT methods has been conducted by dividing the XC functionals in macro-classes, reported roughly according to Jacob's ladder, as GGA, vdW-inclusive functionals, meta-GGA, and hybrid functionals. Also HF was added to the benchmark. The HF, GGA, and hybrid dispersion-less functionals have been corrected through the D3/D4 correction, while the Tkatchenko-Scheffler (TS) and the Many-Body-Dispersion (MBD) corrections have also been taken into account for PBE and PBE0.

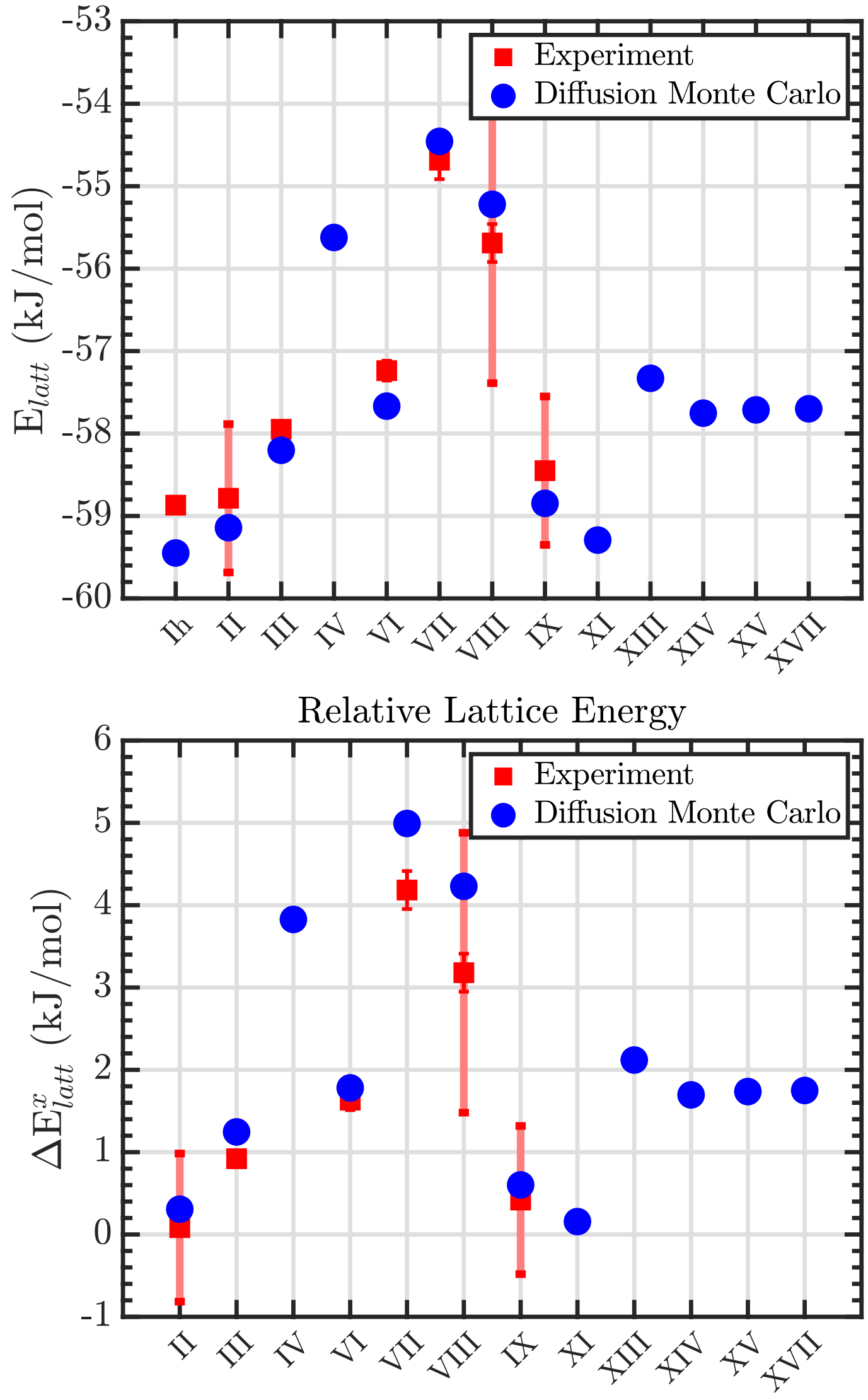

FIG. 2. Performance of DMC for the 13 ice polymorphs considered on the absolute (top) and relative (bottom) lattice energy, compared with available experimental data[25]. The error on the experimental estimates due to the correction for the zero point energy, estimated in the SI, is reported with a shaded error bar for ice II, VIII, and IX. Energies are given kJ/mol.

| | Absolute Lattice Energy | | Relative Lattice Energy | |
|---|---|---|---|---|
| Polymorph | EXP | DMC | EXP | DMC |
| Ih | -58.87(1) | -59.45(7) | – | – |
| II | -58.78(1) | -59.14(7) | 0.084(4) | 0.31(10) |
| III | -57.95(5) | -58.20(7) | 0.92(1) | 1.25(10) |
| IV | n.a. | -55.62(7) | n.a. | 3.83(10) |
| VI | -57.24(12) | -57.67(7) | 1.63(3) | 1.78(10) |
| VII | -54.68(23) | -54.46(7) | 4.18(6) | 4.99(10) |
| VIII | -55.69(23) | -55.22(8) | 3.18(6) | 4.23(10) |
| IX | -58.45(8) | -58.85(7) | 0.42(2) | 0.60(10) |
| XI | n.a. | -59.29(8) | n.a. | 0.15(10) |
| XIII | n.a. | -57.33(7) | n.a. | 2.12(10) |
| XIV | n.a. | -57.75(7) | n.a. | 1.70(10) |
| XV | n.a. | -57.71(7) | n.a. | 1.74(10) |
| XVII | n.a. | -57.70(8) | n.a. | 1.75(10) |

TABLE I. Experimental (Ref. 25) and DMC values (errors given in parentheses) for the absolute and relative lattice energies. Energies are in kJ/mol. [n.a. ≡ not available].

The absolute lattice energies are reported in Table II (we also include LDA lattice energies for completeness, but we leave them out of the general discussion since they are known to be unreliable). The performance of each functional is evaluated as the Mean Absolute Error (MAE) with respect to the reference DMC value, graphically reported both for the absolute and relative lattice energies in Fig. 3 to allow for easier comparisons. In particular, in the following analysis, XC functionals are generally classified as 'good' if their MAE is $\lesssim 4$ kJ/mol ($\lesssim 2$ kJ/mol) for the absolute (relative) lattice energy.

Overall, as expected the worst performance on the absolute lattice energy is obtained at the HF level, with significant improvement ($> 50\%$) gained thanks to the dispersion correction. Except for PBE, the performance of all functionals is improved by the a posteriori vdW correction. Interestingly, the performance on the relative lattice energies are significantly different. Remarkably, the HF method performs better than several DFT functionals, but its performance is significantly worsened by both D3 and D4 corrections. Unreliable results are generally achieved by both PBE and PBE0 regardless of the D3/D4 corrections. Overall, the only macro-classes where all the functionals achieve general good performance (on both the absolute and relative lattice energies) are the meta-GGAs and the vdW-inclusive functionals. As noted before, the energy differences between the considered polymorphs vary in a small range of 5 kJ/mol. For this reason, even if a functional generally achieves small errors in reproducing the reference DMC values for the absolute lattice energy, a fundamental condition to achieve good results on the relative lattice energy is a constant error among all the polymorphs, so that an error cancellation serves to yield the desired performance.

The best performing functionals for each XC macro-class are further analysed in Fig. 4. In the top panel we focus on the absolute lattice energy. Among the considered meta-GGAs, only RSCAN achieves very good results (MAE $\sim 0.9$ kJ/mol), whilst SCAN and R2SCAN have a MAE $\sim$ 4 kJ/mol and $\sim 2.5$ kJ/mol, respectively. Several properties of liquid water have been shown to be correctly predicted by SCAN[14,79,80]. Its performance on a subset of 6 proton-ordered polymorphs (ice IX, II, XIII, XIV, XV, and VIII), has been previously analysed by Sun *et al.*[22]. They showed that SCAN reproduces relative lattice energies better than PBE, PBE0, and PBE0+TS. These considerations are confirmed by our dataset, however, we find several other XC functionals that achieve better performance. Good results are achieved by revPBE-D3 (almost equivalent with or without the Axirold-Teller-Muto correction) and vdW-DF2, with their MAE being lower than 1.4 kJ/mol. However, as evident in both Fig. 4 and Table

II, their error with respect to the DMC reference oscillates among all the polymorphs, leading to worse performance on the stability with respect to hexagonal ice. In the bottom panel of Fig. 4 we focus on the best performing functionals on the relative lattice energies, i.e. revPBE-D3 (GGA), optB86b-vdW (vdW-inclusive), SCAN+rVV10 (meta-GGA), and B3LYP-D3$^{atm}$ (hybrid). Except for revPBE-D3, these functionals generally achieve poor performance when predicting the stability with respect to the gas phase, with a MAE of $\sim 5$ kJ/mol for B3LYP-D3$^{atm}$ and even greater for optB86b-vdW ($\sim 7$ kJ/mol) and SCAN+rVV10 ($\sim 9$ kJ/mol). However, as suggested before, this error behaves like a constant offset with respect to DMC, with the error cancellation allowing for optimal predictions on the relative lattice energy.

Note also that the performance of B3LYP-D3, B3LYP-D3$^{atm}$, optB86b-vdW, and SCAN+rVV10 on the relative lattice energies are all equivalent within the statistical error of the DMC estimates (MAE between 0.5 and 0.7 kJ/mol). However, hybrid functional calculations require computational resources hundreds of times greater than meta-GGA and vdW-inclusive methods (and a denser **k**-point grid is necessary to achieve convergence). Therefore, we suggest the latter to obtain optimal results at a reasonable cost.

Remarkably, the DMC reference values allow us to qualitatively understand why ice III is only meta-stable in the recent computational phase diagrams[14,15]. Since we did not include ice V in our benchmark, we consider its ordered counterpart ice XIII in the following analysis. Neglecting the temperature contribution and the zero point motion, and considering the zero pressure volumes, the transition pressure can be crudely estimated as

$$p_{\text{tr}} = -\frac{\Delta E^{\text{DMC}}_{\text{XIII-III}}}{\Delta V^{p=0}_{\text{XIII-III}}} \sim 0.48 \text{ GPa}. \tag{3}$$

Therefore, even in this approximation our DMC data allow for the prediction of a reasonable transition pressure[11]. Assuming that the difference $\Delta V^0_{\text{XIII-III}}$ does not change significantly with the XC functional, then the condition that a functional needs to satisfy to predict the stability of ice III is $\Delta E^{\text{XC}}_{\text{XIII-III}} \sim \Delta E^{\text{DMC}}_{\text{XIII-III}}$. As can be computed from Table II, none of the functionals considered in Refs. 14 and 15, i.e. SCAN, PBE0+D3, revPBE0+D3, and B3LYP+D3 satisfies this condition. Details of the used approximation are reported in the SI, together with the performance of each XC functional on the transition pressure. Interestingly, none of the best performing functionals identified in our analysis captures this stability difference either.

Based on our benchmarks, it is clear that care should be taken when choosing which functional to use when computing a specific water-ice property. This conclusion is consistent with the recent work of Kapil *et al.*[81], where it was shown that two different functionals, revPBE0-D3 and B3LYP-D3, perform better in capturing the vibrational spectrum of liquid water and hexagonal ice, respectively. From another perspective, despite the large amount of currently available DFT methods, and the optimal (or sub-optimal) performance achieved by several of them, it is evident that there is still room for improvement, especially if we do not want to rely on error cancellation to reach the desired target.

Note also that, despite the large set of XC functionals tested, there are certainly other DFT methods that could be considered. In particular, the recently proposed Density-Corrected SCAN (DC-SCAN)[82,83] looks very promising, while the hybrid functional SCAN0[84] has been tested on liquid water showing better results than SCAN. Moving forward it would be interesting to test these functionals, as well as others left outside this benchmark (e.g. numerous meta-GGAs or double hybrid functionals), on the `DMC-ICE13` set.

## IV. CONCLUSION

We defined a set of thirteen ice polymorphs, including both hydrogen ordered and disordered phases, and ranging from low- to high-pressure phases. We computed absolute and relative lattice energies using DMC, which were shown to be always in excellent agreement (average error of $\sim 0.5$ kJ/mol) with available experimental data. Furthermore, our dataset allows us to qualitatively understand a discrepancy between the experimental and computational phase diagram, highlighting the significance of computing high-accuracy reference data. For polymorphs for which experimental lattice energies are not yet available, our DMC values serve as a useful reference against which new methods can be tested. Indeed, the fact that we can now compute 13 DMC lattice energies in a fairly rapid timescale underlines how DMC has become a powerful reference method for molecular crystals and condensed phase simulations with fairly large unit cells.

Here, we tested a broad range of DFT-XC functionals and, as seen in previous benchmarks, the performance of GGA and hybrid functionals generally improves when a dispersion-correction is taken into account. Of the schemes considered, unreliable performance on the absolute lattice energies are achieved by HF (MAE $> 30$ kJ/mol), revPBE and revPBE0 (MAE $\sim 20$ kJ/mol). However, significant improvement is obtained with D3/D4 corrections for revPBE and revPBE0 (MAE $< 5$ kJ/mol). Almost equivalent performance is achieved by vdW-DF, PBE0+D, and B3LYP+D. The vdW-inclusive functionals generally offer good performance, this being particularly true for the relative lattice energies (MAE $<$ 2 kJ/mol). Interestingly, the performance of an XC functional on the absolute and relative lattice energies can be significantly different, the most evident cases being HF, PBE, RSCAN, and PBE0. Indeed, it is clear from this work that ice phases define a challenging set for electronic structure methods, and that improvements are still needed to capture all its features with a single functional. Overall, our analysis suggests that revPBE-D3 and RSCAN (MAE $\sim 0.9$ kJ/mol) are the most accurate functionals for the absolute lattice energies, while optB86b-vdW and SCAN+rVV10 (MAE $\sim 0.5$ kJ/mol) are the best options - considering both accuracy and computational cost - for the relative lattice energies. Previous comparisons between water clusters and ice phases[21] suggested that these conclusions should be transferable to the liquid state, as well as to solid-water interfaces, where detailed benchmarks

| Method | Ih | II | III | IV | VI | VII | VIII | IX | XI | XIII | XIV | XV | XVII | MAE |
|---|---|---|---|---|---|---|---|---|---|---|---|---|---|---|
| DMC | -59.45 | -59.14 | -58.20 | -55.62 | -57.67 | -54.46 | -55.22 | -58.85 | -59.29 | -57.33 | -57.75 | -57.71 | -57.70 | |
| B3LYP-D4 | -63.03 | -61.11 | -60.42 | -58.10 | -58.69 | -53.78 | -54.98 | -61.33 | -63.20 | -60.18 | -59.60 | -58.65 | -62.05 | 2.20 |
| B3LYP-D3(BJ)$^{atm}$ | -63.63 | -61.90 | -61.11 | -58.90 | -59.60 | -54.96 | -56.21 | -62.03 | -63.80 | -61.00 | -60.43 | -59.56 | -62.68 | 2.88 |
| B3LYP-D3(BJ) | -64.04 | -62.52 | -61.65 | -59.57 | -60.37 | -55.86 | -57.09 | -62.62 | -64.23 | -61.68 | -61.17 | -60.31 | -63.01 | 3.52 |
| B3LYP-D3$^{atm}$ | -64.14 | -64.02 | -62.10 | -60.89 | -62.24 | -59.89 | -60.95 | -63.39 | -64.28 | -62.99 | -62.61 | -62.16 | -63.24 | 4.96 |
| B3LYP-D3 | -64.55 | -64.65 | -62.64 | -61.56 | -63.01 | -60.78 | -61.84 | -63.98 | -64.71 | -63.67 | -63.34 | -62.91 | -63.57 | 5.60 |
| B3LYP | -52.71 | -46.96 | -48.22 | -43.85 | -42.88 | -35.97 | -37.36 | -48.49 | -52.73 | -45.72 | -44.46 | -43.03 | -52.09 | 11.84 |
| revPBE0-D4 | -56.31 | -54.00 | -53.65 | -51.17 | -51.58 | -46.87 | -48.09 | -54.41 | -56.42 | -53.05 | -52.44 | -51.62 | -55.27 | 4.89 |
| revPBE0-D3(BJ)$^{atm}$ | -56.94 | -55.02 | -54.50 | -52.22 | -52.84 | -48.38 | -49.66 | -55.32 | -57.06 | -54.15 | -53.58 | -52.86 | -55.89 | 3.84 |
| revPBE0-D3(BJ) | -57.35 | -55.65 | -55.04 | -52.89 | -53.60 | -49.28 | -50.55 | -55.91 | -57.48 | -54.83 | -54.32 | -53.61 | -56.22 | 3.21 |
| revPBE0-D3$^{atm}$ | -57.03 | -56.96 | -54.81 | -53.80 | -55.43 | -54.71 | -55.78 | -56.03 | -57.09 | -55.65 | -55.34 | -55.40 | -56.11 | 1.99 |
| revPBE0-D3 | -57.45 | -57.58 | -55.35 | -54.47 | -56.20 | -55.61 | -56.67 | -56.61 | -57.51 | -56.33 | -56.08 | -56.15 | -56.44 | 1.63 |
| revPBE0 | -44.05 | -37.58 | -39.51 | -34.72 | -33.31 | -25.88 | -27.35 | -39.63 | -44.00 | -36.41 | -35.03 | -33.58 | -43.29 | 21.08 |
| PBE0-MBD | -65.15 | -63.08 | -62.56 | -60.11 | -60.74 | -55.92 | -57.14 | -63.40 | -65.42 | -62.18 | -61.54 | -60.64 | -64.11 | 4.12 |
| PBE0-TS | -64.67 | -64.14 | -62.66 | -61.30 | -62.52 | -41.88 | -58.46 | -63.76 | -64.95 | -63.44 | -63.13 | -62.43 | -63.62 | 5.67 |
| PBE0-D4 | -64.38 | -61.43 | -61.48 | -58.49 | -58.75 | -53.32 | -54.56 | -62.12 | -64.63 | -60.48 | -59.74 | -58.70 | -63.34 | 2.82 |
| PBE0-D3(BJ)$^{atm}$ | -64.68 | -61.80 | -61.81 | -58.87 | -59.17 | -53.88 | -55.14 | -62.45 | -64.93 | -60.86 | -60.13 | -59.12 | -63.66 | 3.03 |
| PBE0-D3(BJ) | -65.09 | -62.42 | -62.35 | -59.54 | -59.94 | -54.78 | -56.02 | -63.03 | -65.35 | -61.54 | -60.86 | -59.87 | -63.99 | 3.57 |
| PBE0-D3$^{atm}$ | -65.80 | -64.77 | -63.44 | -61.18 | -61.90 | -57.86 | -58.99 | -64.35 | -66.05 | -63.22 | -62.62 | -61.82 | -64.81 | 5.26 |
| PBE0-D3 | -66.21 | -64.77 | -63.98 | -61.85 | -62.67 | -58.75 | -59.88 | -64.94 | -66.47 | -63.91 | -63.35 | -62.57 | -65.14 | 5.85 |
| PBE0 | -57.47 | -52.12 | -53.27 | -49.05 | -48.39 | -41.88 | -43.23 | -53.49 | -57.63 | -50.88 | -49.74 | -48.45 | -56.77 | 6.62 |
| SCAN+rVV10 | -68.26 | -67.83 | -66.33 | -65.17 | -66.27 | -64.07 | -65.53 | -67.38 | -68.52 | -67.14 | -66.90 | -66.61 | -67.29 | 9.15 |
| R2SCAN | -62.83 | -61.34 | -60.39 | -58.60 | -59.53 | -56.39 | -57.61 | -61.00 | -63.08 | -60.40 | -60.04 | -59.43 | -62.10 | 2.64 |
| RSCAN | -61.40 | -59.38 | -58.67 | -56.51 | -57.35 | -53.83 | -55.04 | -59.15 | -61.68 | -58.35 | -57.90 | -57.21 | -60.75 | 0.93 |
| SCAN | -64.42 | -62.84 | -61.95 | -60.05 | -60.58 | -57.58 | -59.09 | -62.77 | -64.80 | -61.95 | -61.46 | -61.00 | -63.69 | 4.14 |
| optB88-vdW | -67.90 | -67.85 | -66.91 | -65.66 | -66.99 | -63.50 | -64.61 | -67.88 | -68.32 | -67.66 | -67.52 | -66.82 | -66.40 | 9.33 |
| optB86b-vdW | -68.69 | -67.89 | -67.47 | -65.74 | -66.83 | -62.84 | -63.93 | -68.22 | -69.18 | -67.68 | -67.43 | -66.63 | -67.21 | 7.05 |
| optPBE-vdW | -65.55 | -65.72 | -64.79 | -63.77 | -64.84 | -61.25 | -62.46 | -65.93 | -65.82 | -65.66 | -65.50 | -64.83 | -63.87 | 9.20 |
| rev-vdW-DF2 | -66.37 | -64.26 | -64.27 | -61.87 | -62.52 | -57.90 | -59.03 | -64.84 | -66.84 | -63.79 | -63.32 | -62.35 | -65.26 | 5.71 |
| vdW-DF2 | -59.43 | -60.16 | -58.40 | -57.93 | -59.18 | -56.49 | -57.80 | -59.87 | -59.43 | -59.92 | -59.84 | -59.30 | -58.06 | 1.34 |
| vdW-DF | -53.59 | -53.78 | -52.86 | -52.03 | -52.74 | -49.13 | -50.47 | -54.19 | -53.59 | -53.81 | -53.60 | -52.96 | -51.84 | 4.91 |
| revPBE-D4 | -59.96 | -56.47 | -57.12 | -54.00 | -53.97 | -48.67 | -49.73 | -57.33 | -60.29 | -55.68 | -54.98 | -53.90 | -58.96 | 2.53 |
| revPBE-D3(BJ)$^{atm}$ | -58.86 | -55.92 | -56.39 | -53.54 | -53.66 | -48.25 | -49.43 | -56.74 | -59.17 | -55.31 | -54.64 | -53.62 | -57.80 | 2.71 |
| revPBE-D3(BJ) | -59.27 | -56.55 | -56.92 | -54.21 | -54.42 | -49.15 | -50.31 | -57.33 | -59.59 | -55.99 | -55.37 | -54.38 | -58.13 | 2.17 |
| revPBE-D3$^{atm}$ | -58.60 | -57.12 | -56.15 | -54.37 | -55.40 | -53.93 | -54.85 | -56.82 | -58.83 | -56.02 | -55.57 | -55.31 | -57.68 | 1.36 |
| revPBE-D3 | -59.01 | -57.75 | -56.69 | -55.04 | -56.16 | -54.83 | -55.74 | -57.41 | -59.25 | -56.71 | -56.30 | -56.07 | -58.00 | 0.91 |
| revPBE | -43.86 | -35.59 | -38.96 | -33.16 | -30.88 | -21.89 | -23.30 | -38.52 | -43.97 | -34.66 | -33.02 | -31.14 | -43.11 | 22.79 |
| PBE-MBD | -70.48 | -67.29 | -67.77 | -64.69 | -64.75 | -58.87 | -59.94 | -68.13 | -70.96 | -66.64 | -65.83 | -64.57 | -69.40 | 8.53 |
| PBE-TS | -69.59 | -67.75 | -67.44 | -65.29 | -65.86 | -58.84 | -60.36 | -68.04 | -70.10 | -67.30 | -66.81 | -65.71 | -68.53 | 8.71 |
| PBE-D4 | -69.62 | -65.55 | -66.60 | -62.97 | -62.71 | -56.21 | -57.35 | -66.75 | -70.10 | -64.82 | -63.96 | -62.58 | -68.61 | 6.88 |
| PBE-D3(BJ)$^{atm}$ | -69.95 | -65.95 | -66.96 | -63.38 | -63.19 | -56.85 | -58.02 | -67.11 | -70.43 | -65.23 | -64.38 | -63.05 | -68.96 | 7.31 |
| PBE-D3(BJ) | -70.36 | -66.58 | -67.49 | -64.05 | -63.96 | -57.75 | -58.90 | -67.69 | -70.85 | -65.92 | -65.11 | -63.80 | -69.29 | 7.95 |
| PBE-D3$^{atm}$ | -70.39 | -67.75 | -67.94 | -65.15 | -65.41 | -59.76 | -60.90 | -68.38 | -70.87 | -67.07 | -66.37 | -65.25 | -69.42 | 8.94 |
| PBE-D3 | -70.80 | -68.37 | -68.47 | -65.82 | -66.17 | -60.66 | -61.79 | -68.97 | -71.29 | -67.76 | -67.11 | -66.00 | -69.75 | 9.58 |
| PBE | -62.23 | -55.54 | -57.85 | -52.85 | -51.58 | -43.79 | -45.05 | -57.56 | -62.60 | -54.54 | -53.24 | -51.56 | -61.52 | 4.48 |
| HF-D4 | -42.89 | -46.50 | -41.70 | -42.92 | -45.69 | -46.31 | -47.99 | -44.36 | -42.18 | -45.14 | -45.47 | -46.08 | -41.82 | 13.03 |
| HF-D3(BJ)$^{atm}$ | -50.53 | -53.57 | -49.36 | -50.04 | -53.03 | -53.70 | -55.15 | -51.58 | -49.97 | -52.28 | -52.77 | -53.25 | -49.53 | 5.67 |
| HF-D3(BJ) | -50.94 | -54.20 | -49.90 | -50.70 | -53.79 | -54.60 | -56.03 | -52.16 | -50.39 | -52.97 | -53.50 | -54.00 | -49.86 | 5.17 |
| HF-D3$^{atm}$ | -39.07 | -44.94 | -38.65 | -41.32 | -44.88 | -47.30 | -49.09 | -41.88 | -38.27 | -43.58 | -44.14 | -45.31 | -38.01 | 14.76 |
| HF-D3 | -39.49 | -45.57 | -39.19 | -41.99 | -45.64 | -48.20 | -49.97 | -42.47 | -38.69 | -44.27 | -44.88 | -46.06 | -38.34 | 14.13 |
| HF | -26.57 | -25.53 | -23.38 | -22.01 | -22.58 | -19.54 | -21.57 | -25.42 | -25.62 | -24.06 | -23.47 | -23.29 | -25.73 | 33.82 |
| LDA | -100.08 | -94.37 | -95.92 | -90.94 | -91.03 | -83.78 | -84.66 | -95.37 | -101.22 | -93.04 | -91.99 | -90.32 | -99.67 | 35.69 |

TABLE II. Performance of each exchange-correlation functional on the absolute lattice energy. Energies are reported in kJ/mol. In order to facilitate reproducibility, in the SI we report the same table with energies in meV, directly comparable to the VASP output.

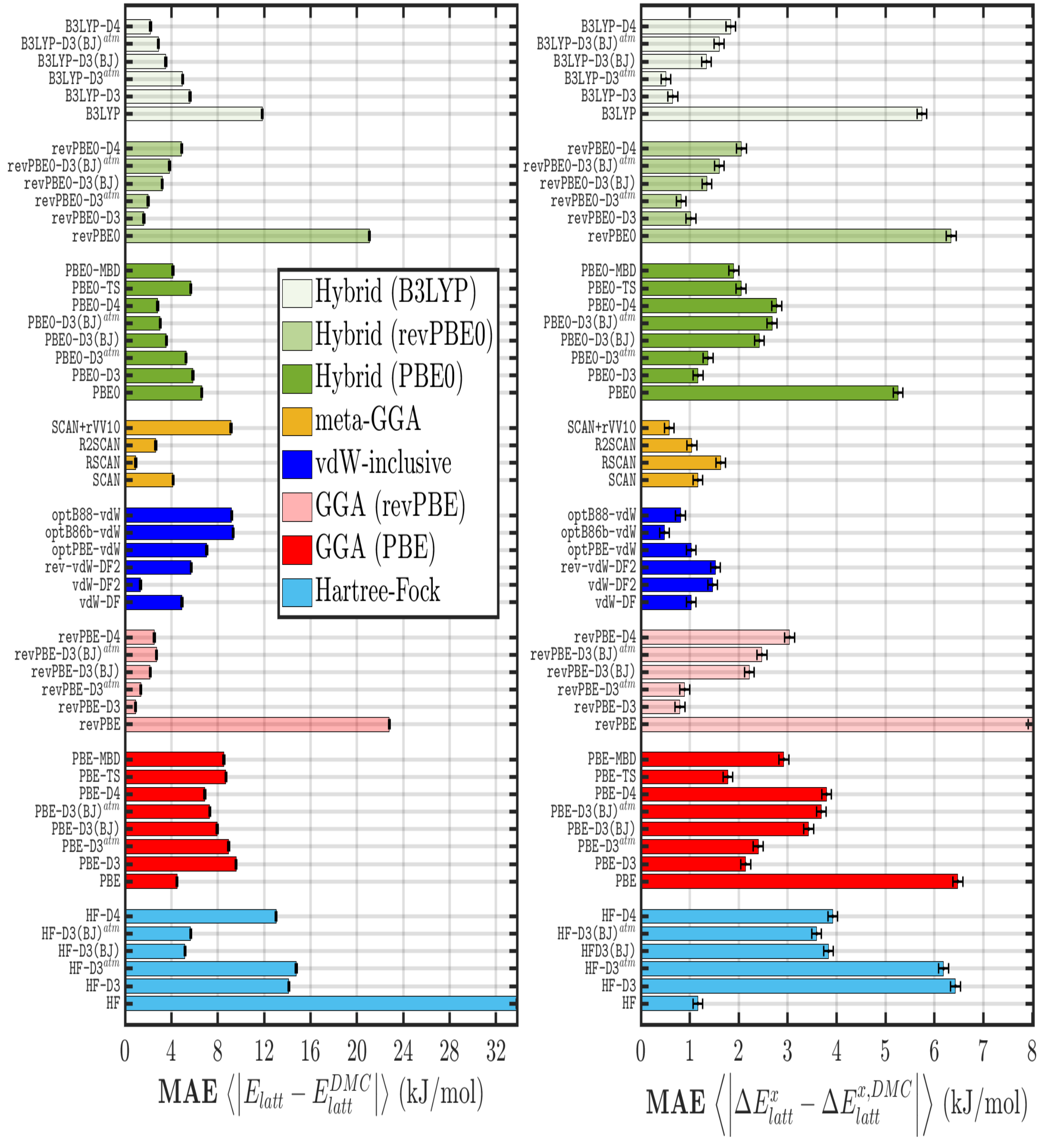

FIG. 3. Outcome of the benchmark, reported as Mean Absolute Error (MAE) with respect to the reference DMC values (black error bars indicate the DMC statistical error), for the absolute (left) and relative (right) lattice energy. The XC functionals have been ordered according to Jacob's ladder as Generalised Gradient Approximation (red), non-local van der Waals inclusive functionals (blue), meta-GGA (ochre), and hybrid (green). Different shades are used within GGA and hybrid methods to differentiate the basis dispersion-less functional. Hartree-Fock (cyan) results are also reported.

are hindered by the prohibitive computational cost.

## ACKNOWLEDGEMENTS

This research used resources of the Oak Ridge Leadership Computing Facility at the Oak Ridge National Labo-

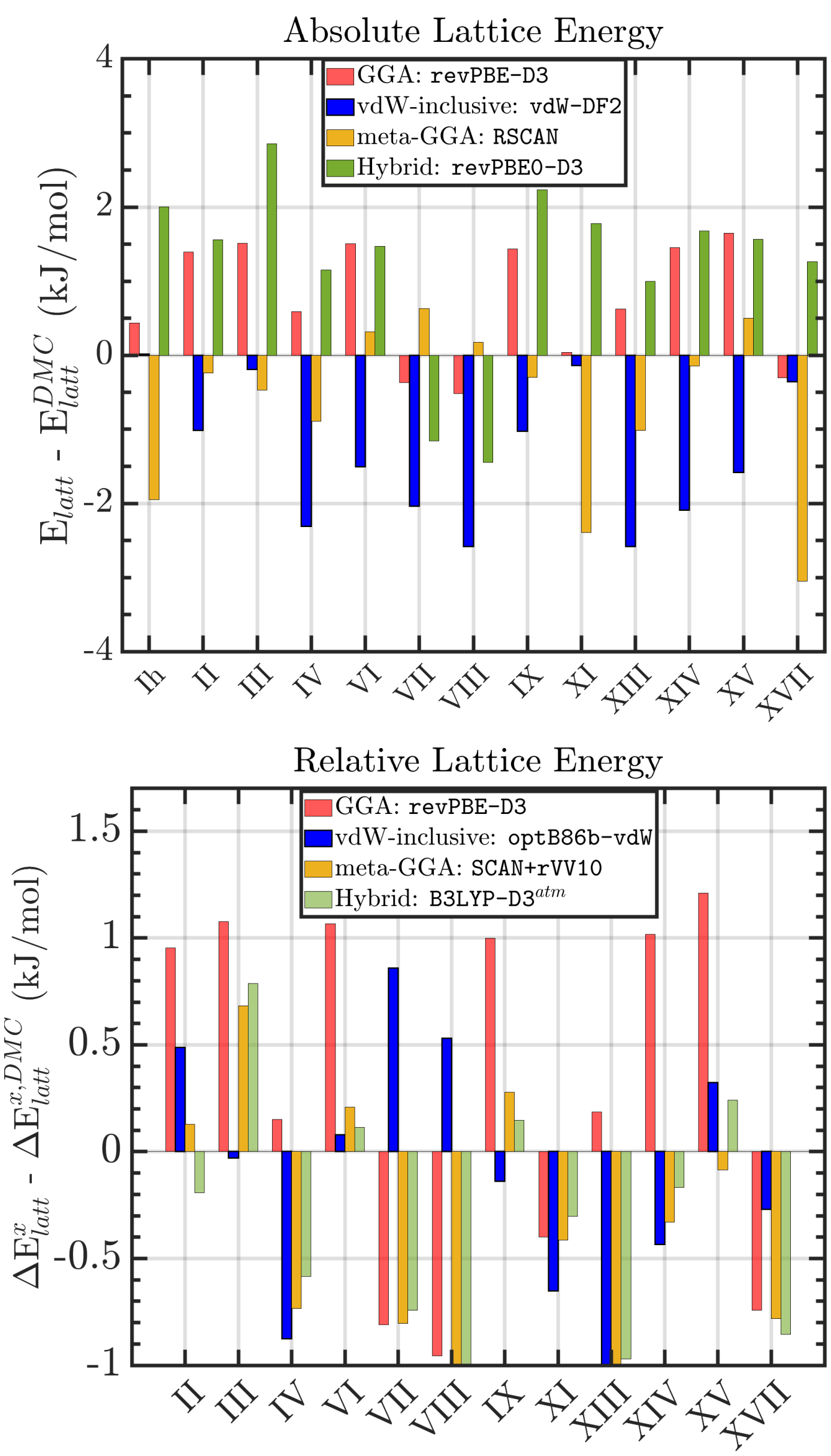

FIG. 4. Best performing DFT functional for each XC macro-class on the absolute (top) and relative (bottom) lattice energy. Each plot reports the difference between the DFT-XC and the reference DMC lattice energy for all the polymorphs in `DMC-ICE13`. Each XC functional bar is coloured according to the colour-map used in Fig. 3.

ratory, which is supported by the Office of Science of the U.S. Department of Energy under Contract No. DE-AC05-00OR22725). Calculations were also performed using the Cambridge Service for Data Driven Discovery (CSD3) operated by the University of Cambridge Research Computing Service (www.csd3.cam.ac.uk), provided by Dell EMC and Intel using Tier-2 funding from the Engineering and Physical Sciences Research Council (capital grant EP/T022159/1 and EP/P020259/1), and DiRAC funding from the Science and Technology Facilities Council (www.dirac.ac.uk). This work also used the ARCHER UK National Supercomputing Service (https://www.archer2.ac.uk), the United Kingdom Car Parrinello (UKCP) consortium (EP/ F036884/1).

# Supporting Information for ‘DMC-ICE13: ambient and high pressure polymorphs of ice from Diffusion Monte Carlo and Density Functional Theory’

Flaviano Della Pia [1], Andrea Zen [2,3], Dario Alfè [2,3,4,5], and Angelos Michaelides [1]

[1] Yusuf Hamied Department of Chemistry, University of Cambridge, Cambridge CB2 1EW, United Kingdom
[2] Dipartimento di Fisica Ettore Pancini, Università di Napoli Federico II, Monte S. Angelo, I-80126 Napoli, Italy
[3] Department of Earth Sciences, University College London, London WC1E 6BT, United Kingdom
[4] Thomas Young Centre, University College London, London WC1E 6BT, United Kingdom
[5] London Centre for Nanotechnology, University College London, London WC1E 6BT, United Kingdom

In section 1 we describe the approximation used to qualitatively understand the discrepancy between the experimental and computational phase diagrams regarding the ice III-V phase transition. In particular, we provide the performance of each considered Density Functional Theory (DFT) exchange-correlation (XC) functional on the transition pressure, as compared to our ‘reference’ value computed with the Diffusion Monte Carlo (DMC) estimates of the lattice energy. In Section 2 we describe how the lattice energy can be obtained from experimental evaluations of the sublimation enthalpy, and we discuss the corresponding uncertainty. Possible errors due to the PBE geometry relaxation of the structures in DMC-ICE13 are discussed in Section 3. In section 4 we conduct a DFT analysis on the effect of hydrogen-disorder on the lattice energy. We provide details regarding the Finite Size Errors (FSE) and the time step convergence in our DMC simulations respectively in Section 5 and 6. In Section 7 we provide the DFT results for all the considered polymorphs and functionals, with energy given in meV to facilitate the comparison with the VASP output. In section 8 we provide both DMC and DFT outputs for the water monomer considered in the computation of the absolute lattice energies. Finally, we provide all the geometries of DMC-ICE13 in VASP POSCAR format in Section 9. A complete directory with all the DFT and DMC input files is provided as supplementary material.

## 1 Discrepancy between experimental and computational phase diagrams

As briefly discussed in the main manuscript, the DMC estimates of the lattice energy allow us to qualitatively understand why ice III is only meta-stable in the recent phase diagrams [1, 2], computed using the DFT functionals SCAN, PBE0+D3, revPBE0+D3, and B3LYP+D3.

The small stability region of ice III is identified by temperatures in the range $\sim [240-260]$ K, and pressures in the range $\sim [0.2-0.5]$ GPa. At fixed temperature - this domain is included between the stability region of ice V and Ih. Since we did not include ice V in our dataset, we consider its ordered counterpart ice XIII in the following analysis. The DMC energy difference between ice XIII and III is

$$\Delta E^{\text{DMC}}_{\text{XIII-III}} = E^{\text{DMC}}_{\text{XIII}} - E^{\text{DMC}}_{\text{III}} \sim 0.87 \pm 0.1 \text{ kJ/mol}. \quad (1)$$

Neglecting the temperature contribution and the zero point motion, and considering the zero pressure volumes, the only additional term in the free energy difference (enthalpy difference) is the pressure dependent term $p\Delta V^0_{\text{XIII-III}}$.

This simple approximation allows for a qualitative prediction of the transition pressure as

$$p_{\text{tr}}^{\text{DMC}} = -\frac{\Delta E^{\text{DMC}}_{\text{XIII-III}}}{\Delta V^{p=0}_{\text{XIII-III}}} \sim 0.48 \text{ GPa}. \tag{2}$$

Remarkably, this is in agreement with the experimental phase diagram [3]. Assuming that the difference $\Delta V^0_{\text{XIII-III}}$ does not change significantly with the XC functional, then the condition that a functional needs to satisfy to predict the stability of ice III is $\Delta E^{XC}_{XIII-III} \sim \Delta E^{DMC}_{XIII-III} \sim 0.87$ kJ/mol. Equivalently, we can estimate the transition pressure predicted by each XC functional as

$$p_{\text{tr}}^{\text{XC}} = -\frac{\Delta E^{\text{XC}}_{\text{XIII-III}}}{\Delta V^{p=0}_{\text{XIII-III}}}. \tag{3}$$

The performance of each XC functional on the energy variation $\Delta E^{\text{XC}}_{XIII-III}$ and on the predicted transition pressure $p_{\text{tr}}^{\text{XC}}$ are reported in Fig. 1. Interestingly, none of the best performing functionals identified in our benchmark obtains reliable results.

Unfortunately, this analysis can not be used to explain why ice VI is less stable in the simulations than in the experiment. In fact, ice XIII and VI are almost degenerate at the DMC level ($\Delta E^{\text{DMC}}_{\text{VI-XIII}} \sim -0.3 \pm 0.1$ kJ/mol). Moreover, the phase transition happens at higher pressures of $\sim 1$ GPa. Accurate estimates of the pressure and temperature dependent terms are necessary to understand the phase transition.

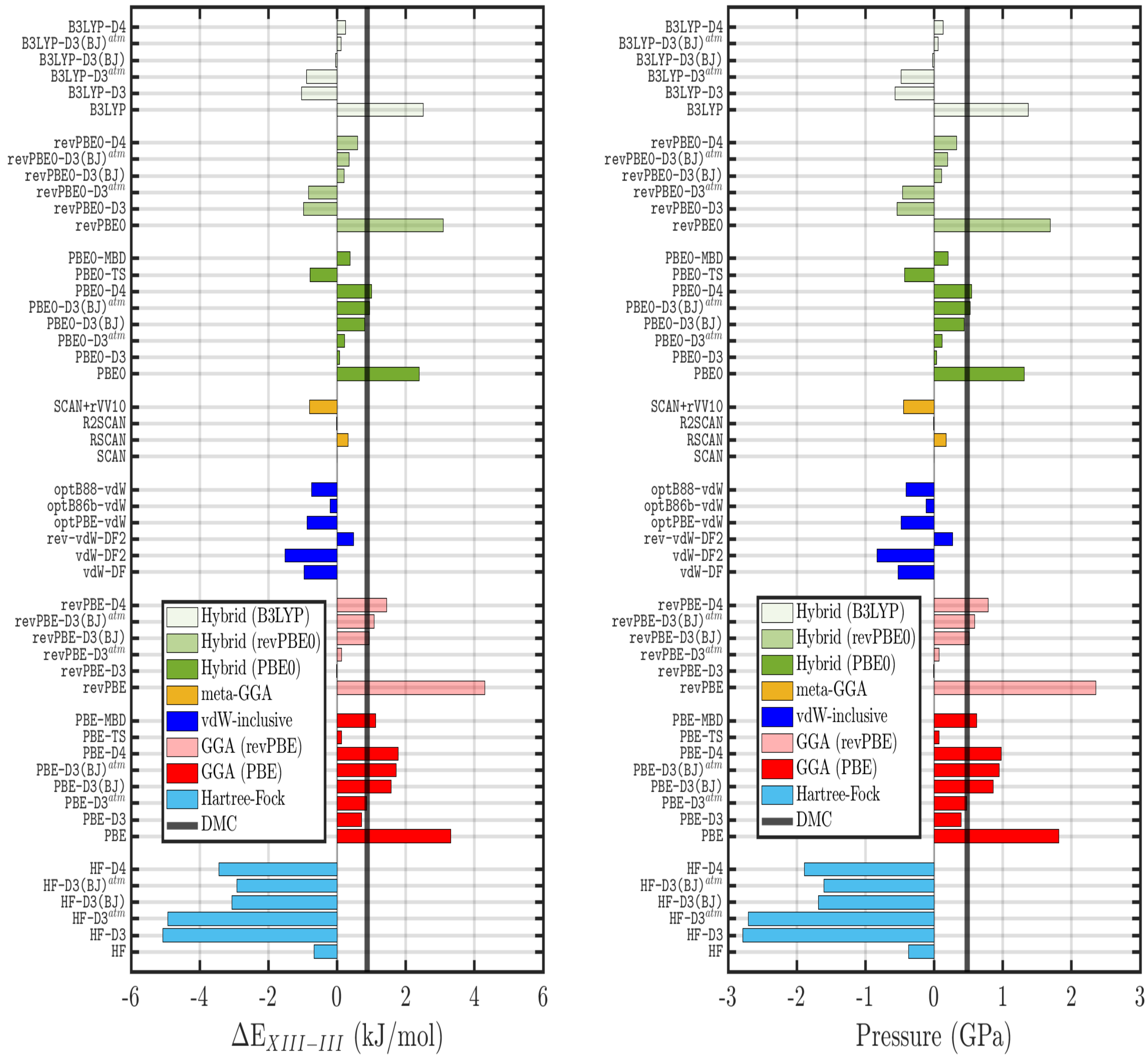

Figure 1: Performance of each XC functional on $\Delta E_{III-XIII}$ (left), as defined in Eq. 1, and the transition pressure (right) between ice XIII/V and III, computed according to Eq. 3. The black vertical line indicates the reference value computed with DMC.

## 2 Experimental estimates of the lattice energies

The available experimental estimates of the internal energy variation of several ice polymorphs (II, III, VI, VII, VIII, IX) with respect to ice Ih, were provided by Whalley [4] in 1984, through measures of sublimation enthalpies, and a subsequent zero temperature and pressure extrapolation. The zero temperature-pressure internal energy difference between ice and the gas phase can be written $\Delta U = E_{\text{latt}} + \Delta\text{ZPE}$, where $E_{\text{latt}}$ is the absolute lattice energy, and $\Delta$ZPE is the difference in the Zero Point Energy (ZPE) between the solid and the gas phase. Therefore, the electronic lattice energy of a generic polymorph 'x' can be estimated from the available data as:

$$E^{x}_{\text{latt}} = E^{\text{Ih}}_{\text{latt}} + \left(\Delta U^{x} - \Delta U^{\text{Ih}}\right) + \left(\Delta ZPE^{\text{Ih}} - \Delta ZPE^{x}\right). \quad (4)$$

The common assumption adopted in previous benchmarks is to neglect the difference in the Zero Point Energy, leading to the experimental values reported in Table 1 of the main manuscript.
Recently, the importance of the ZPE for crystalline ice phases has been theoretically investigated by Rasti and Meyer [5], who provided DFT estimates of the lattice energies, with and without ZPE, for six ice polymorphs (Ih, VIII, IX, XIII, XIV, XV), and with five different XC functionals (PBE, PBE+TS, PBE+MBD, SCAN, SCAN+rVV10).
The data reported in this manuscript allow us to estimate the difference $\left(\Delta ZPE^{\text{Ih}} - \Delta ZPE^{x}\right)$, reported for simplicity in Table 1.

| | PBE | PBE+TS | PBE+MBD | SCAN | SCAN+rVV10 |
|---|---|---|---|---|---|
| II | -2 | 2 | 3 | 7 | 4 |
| VIII | -4 | -2 | -17 | 4 | 17 |
| IX | -1 | -1 | -5 | 4 | 4 |
| XIII | -4 | -4 | -2 | 10 | 5 |
| XIV | -3 | 1 | -4 | 10 | 2 |
| XV | -4 | 2 | -7 | 7 | 2 |

Table 1: Difference $\left(\Delta ZPE^{\text{Ih}} - \Delta ZPE^{x}\right)$ obtained from data in Ref. [5]. Energies are in meV.

Without computing the ZPE at a higher level of theory it is not possible to establish which functional gives the most accurate result. However, it is clear that the assumption $\Delta ZPE^{\text{Ih}} = \Delta ZPE^{x}$ could lead to an underestimation of the experimental error bar, that we estimate to be of the order of $1-2$ kJ/mol ($\sim 10-20$ meV). In particular, based on the reported data, we estimated the experimental error bar to be of $\sim 0.9$ kJ/mol ($\sim 9$ meV) for ice II and IX, while it is $\sim 1.7$ kJ/mol ($\sim 17$ meV) for ice VIII.

## 3 Effect of relaxation with other XC functionals

As described in Sec. 2 of the main manuscript, the `DMC-ICE13` dataset is constructed optimising the geometries of 13 ice polymorphs at the DFT-PBE level. This choice being motivated by excellent agreement between experimental and previous DMC estimates of the ice polymorphs lattice energies on PBE geometries [6, 7]. In this section, we briefly discuss the potential error due to this approximation. In particular, we considered two ice polymorphs, ice Ih and VIII, representative of the low and high pressure region of the water phase diagram, respectively. The structures of these polymorphs were optimised with the XC functional revPBE-D3. This functional achieves the best performances on the absolute lattice energies (MAE $\sim$ 1 kJ/mol) on the `DMC-ICE13` dataset, and was also shown to give good performances in the equilibrium volumes benchmark on the `ICE10` dataset [8]. In Table 2 we describe the effect of the relaxation with revPBE-D3 on the two considered polymorphs, reporting the equilibrium volumes, the revPBE-D3 and the DMC lattice energies on both the PBE and revPBE-D3 geometries. For ice Ih, the revPBE-D3 optimisation results in a $\sim 3\%$ volume variation with respect to PBE, while we find a minimal 0.4% variation in the DFT (revPBE-D3) lattice energies. This estimate is in agreement with the DMC reference value. Relaxation effects are slightly greater for the high-pressure ice VIII, with a percentage volume variation of $\sim 10\%$, and a $\sim 2\%$ variation of the revPBE-D3 lattice energy. This variation is also confirmed by the DMC simulation.
Since we have considered both a low and a high pressure stable polymorph, we believe this analysis to be representative of the entire dataset. Therefore, we expect possible different (more accurate) XC-geometries to lead to a $\sim 1$ kJ/mol variation in the lattice energy. We leave for the future a more extensive analysis and benchmark on the issue of equilibrium volumes.

| Ice Ih | | | |
|---|---|---|---|
| Geometry | Volume | Lattice Energy (revPBE-D3) | Lattice Energy (DMC) |
| PBE | 30.78 | -58.71 | -59.45(7) |
| revPBE-D3 | 31.64 | -58.95 | -60.13(7) |

| Ice VIII | | | |
|---|---|---|---|
| Geometry | Volume | Lattice Energy (revPBE-D3) | Lattice Energy (DMC) |
| PBE | 20.74 | -55.45 | -55.22(8) |
| revPBE-D3 | 18.72 | -56.56 | -55.61(10) |

Table 2: Effect of the relaxation with the revPBE-D3 functional on the lattice energies of `DMC-ICE13`. Equilibrium volumes per molecule are in $\text{Å}^3$ and energies are in kJ/mol.

## 4 DFT analysis of hydrogen-disordered polymorphs

The `DMC-ICE13` dataset contains both hydrogen-ordered and disordered polymorphs. In principle, when computing the lattice energy of a hydrogen-disordered polymorph several structures should be considered, leading to slightly different lattice energies. To address this problem, we considered five hydrogen-disordered polymorphs, ice III, IV, V, VI, and XII (note that only ice III, IV and VI are actually included in `DMC-ICE13`). For each polymorph, we considered five different structures taken from Ref. [9], i.e. five different hydrogen arrangements. First, we optimised each structure at the PBE level to be consistent with the procedure adopted in all our simulations. Then, we computed the lattice energy for each structure at the DFT level, using the revPBE-D3 functional, expected to give results close to reference DMC according to our benchmark.
In Fig. 2 we report the energy differences among the considered structures with respect to a fixed arrangement, for all the analysed polymorphs. Our results suggest that very expensive DMC simulations would be needed to highlight lattice energy differences among different hydrogen arrangements in hydrogen-disordered polymorphs. Therefore, both the DMC and the DFT analyses conducted on the `DMC-ICE13` should not be affected by having considered only one possible configuration for the hydrogen-disordered structures.

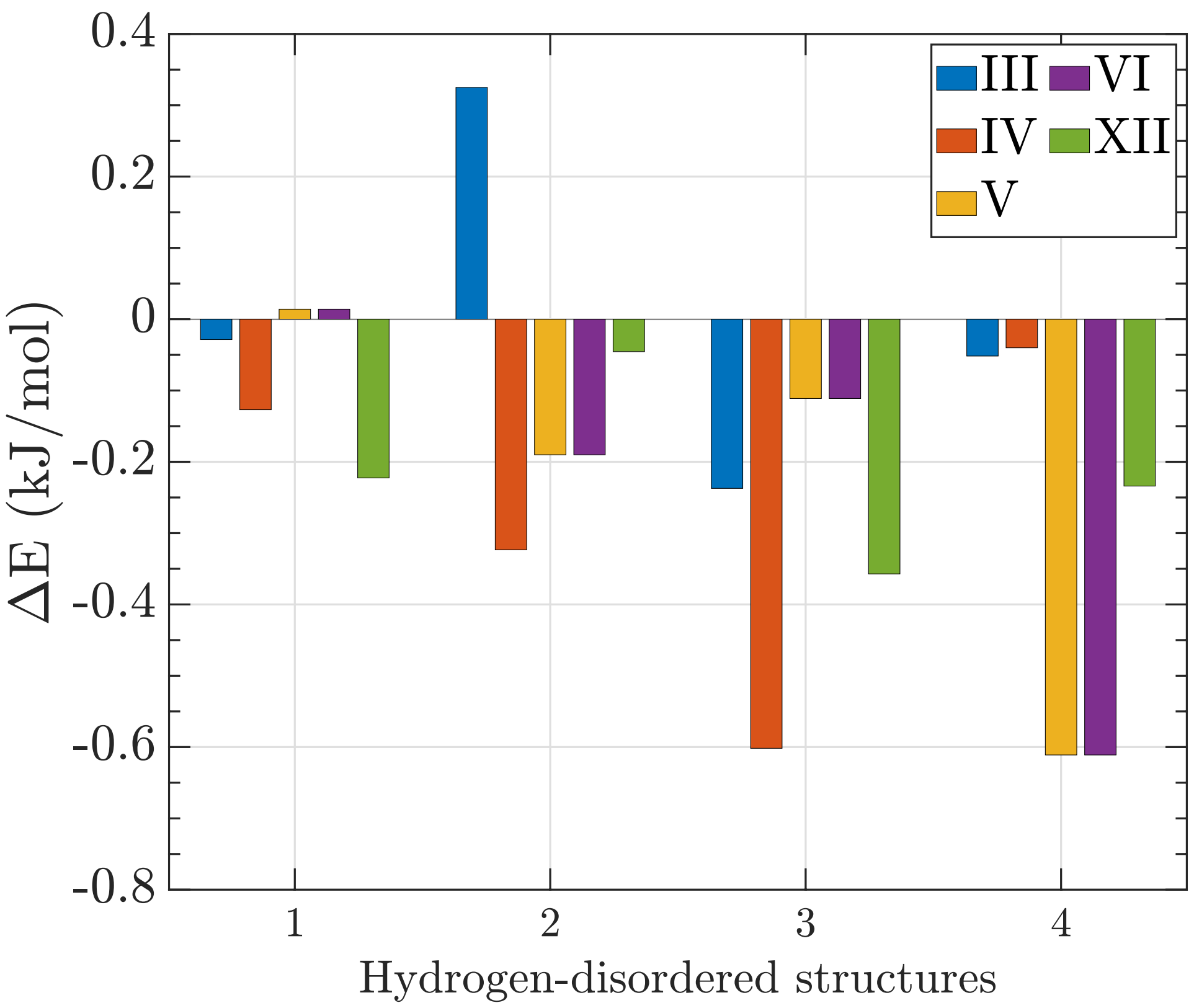

Figure 2: Lattice energy differences for different hydrogen arrangements in ice III, IV, V, VI, and XII. For each polymorph, we considered five different hydrogen distributions and reported the difference in the lattice energy with respect to a fixed configuration.

# 5 Finite Size Errors in DMC simulations

In this section we provide an insight into the FSE of our DMC simulations. The computation of $E_{\text{crys}}$ (as defined in Eq. 1 of the main manuscript) involves the use of periodic boundary conditions, that in DMC can result in significant FSE. However, several corrections schemes to reduce FSE are available in DMC [10, 11, 12, 13, 14]. These methods have been tested extensively on molecular crystals in Ref. [7]: the Model Periodic Coulomb (MPC) [12, 13, 14] correction provided the best results, and therefore it has been used in all our simulations. A second and smaller source of FSE in DMC is due to the use of single-particle orbitals obtained from a DFT calculation on a single point in the Brillouin zone and is called the Independent Particle FSE (IPFSE). Indicating with $E_{\text{crys}}^{\text{DFT},\infty}$ the converged energy per molecule in the crystal, obtained by considering Monkhorst-Pack **k**-points grid of increasing size, then a good estimation of the IPFSE for a $l \times m \times n$ supercell is given by IPFSE $= E_{\text{crys}}^{\text{DFT},l\times m\times n} - E_{\text{crys}}^{\text{DFT},\infty}$. Finally, the DMC simulations with PBC have been performed using Twist Averaging Boundary Condition (TABC) [15].

In Table 3 we report the difference between the MPC and the EWALD estimates of the lattice energies, the IPFSE, and the number of twists for all the polymorphs in `DMC-ICE13`, specifying the number of molecules in the simulated cell. An insight on the effect of a bigger simulation cell on the MPC and EWALD estimates of the lattice energies is reported in Tables 4, 5, and 6 for ice VIII, XI, and XVII. Increasing the size of the simulated cell significantly reduces the difference between the MPC and EWALD estimates, with a variation of $\sim$ 16 kJ/mol ($\sim$ 50%) for ice XI and $\sim$ 35 kJ/mol ($\sim$ 75%) for ice VIII and XVII. However, the correction to the MPC value is much smaller ($\sim$ 1.5 kJ/mol in all the cases), suggesting that the MPC scheme captures most of the FSE even with a small simulated cell.

| Polymorph | # MOL | # TWISTS | MPC-EWALD | IPFSE |
|---|---|---|---|---|
| Ih | 12 | 4 | 23.4 | -0.004 |
| II | 12 | 4 | 28.4 | -0.02 |
| III | 12 | 4 | 25.1 | -0.02 |
| IV | 16 | 4 | 21.6 | -0.002 |
| VI | 10 | 14 | 36.1 | -0.001 |
| VII | 16 | 4 | 15.9 | -0.01 |
| VIII | 32 | 14 | 12.41 | -0.0001 |
| IX | 12 | 4 | 26.7 | -0.01 |
| XI | 16 | 4 | 17.7 | -0.18 |
| XIII | 28 | 4 | 12.75 | -0.001 |
| XIV | 12 | 14 | 27.9 | -0.007 |
| XV | 10 | 14 | 35.7 | -0.008 |
| XVII | 24 | 4 | 10.8 | -0.003 |

Table 3: Analysis of the FSE for each polymorph in `DMC-ICE13`. We report the number of molecules in the simulated cell, the number of twists used for the TABC, the difference between the MPC and the EWALD estimates of the lattice energy, and the IPFSE. Energies are in kJ/mol.

| unit cell | # MOL | MPC-EWALD | MPC |
|---|---|---|---|
| 1 ×1 × 1 | 8 | 47.56 | -56.15(7) |
| 2 ×2 × 1 | 32 | 12.41 | -55.22(8) |

Table 4: Insight on the effect of the simulated cell size on the FSE error for ice VIII. We report the difference between the MPC and EWALD schemes, as well as the MPC estimate of the lattice energy, for a small ($1 \times 1 \times 1$) and a large ($2 \times 2 \times 1$) simulated cell.

| unit cell | # MOL | MPC-EWALD | MPC |
|---|---|---|---|
| 1 ×1 × 1 | 8 | 33.66 | -59.29(7) |
| 2 ×1 × 1 | 16 | 17.7 | -60.55(7) |

Table 5: Insight on the effect of the simulated cell size on the FSE error for ice XI. We report the difference between the MPC and EWALD schemes, as well as the MPC estimate of the lattice energy, for a small ($1 \times 1 \times 1$) and a large ($2 \times 1 \times 1$) simulated cell.

| unit cell | # MOL | MPC-EWALD | MPC |
|---|---|---|---|
| 1 ×1 × 1 | 6 | 44.4 | -56.30(8) |
| 2 ×2 × 1 | 24 | 10.8 | -57.70(8) |

Table 6: Insight on the effect of the simulated cell size on the FSE error for ice XVII. We report the difference between the MPC and EWALD schemes, as well as the MPC estimate of the lattice energy, for a small ($1 \times 1 \times 1$) and a large ($2 \times 2 \times 1$) simulated cell.

# 6 Time step convergence in DMC simulations

Table 7 shows the absolute lattice energy $E_{\mathrm{latt}}$, corrected for the FSE, as a function of the DMC time step for all the considered polymorphs.

All the lattice energies have been computed using TABC, therefore each value in Table 7, and the respective error bar, is obtained as an average over the different twists in the Brillouin zone.

As already reported in the analysis on the molecular crystals in Ref. [7], the MPC correction and the ZSGMA algorithm lead in general to a quite small time step dependence, that is less than 0.5 kJ/mol for $\tau \leq 0.01$ au (for all the polymorphs but ice II). However, the time step dependence is in general different among the polymorphs, thus all the lattice energies reported in the main paper have been computed with a time step of 0.003 au. Additional tests with $\tau = 0.001$ au have been conducted to check the convergence of our estimates for ice Ih, II, VIII, IX, and XIV.

Note here that the energy scales in the time step convergence can be extremely different among different polymorphs. For instance, when $\tau$ goes from 0.05 to 0.003 au, the lattice energy of ice VII varies in a range of only 0.1 kJ/mol. By contrast, the lattice energy of ice II varies in a range of 10 kJ/mol. To showcase these differences, we report all the time step convergence behaviours on a single scale in Fig. 3.

**Time step convergence of the absolute lattice energy**

| $\tau$ | 0.001 | 0.003 | 0.01 | 0.05 |
|---|---|---|---|---|
| Ih | -59.44(14) | -59.45(7) | -59.63(6) | -60.95(6) |
| II | -58.8(3) | -59.14(7) | -60.40(6) | -70.13(6) |
| III | * | -58.20(7) | -58.13(6) | -57.92(6) |
| IV | * | -55.62(7) | -55.78(8) | -57.22(7) |
| VI | * | -57.67(7) | -57.44(6) | -56.76(6) |
| VII | * | -54.46(7) | -54.51(6) | -54.45(7) |
| VIII | -55.1(3) | -55.22(8) | -54.99(8) | -54.19(7) |
| IX | -58.79(14) | -58.85(7) | -58.77(6) | -58.36(6) |
| XI | * | -59.29(8) | -58.91(11) | -58.35(7) |
| XIII | * | -57.33(7) | -57.30(7) | -57.40(6) |
| XIV | -57.79(14) | -57.75(7) | -57.52(7) | -56.03(7) |
| XV | * | -57.71(7) | -57.63(6) | -56.84(6) |
| XVII | * | -57.70 (8) | -57.70 (7) | -55.92 (11) |

Table 7: Time step dependence of the absolute lattice energies for all the polymorphs in `DMC-ICE13`. Energies are in kJ/mol, while the time step is given in au. [* $\equiv$ not computed].

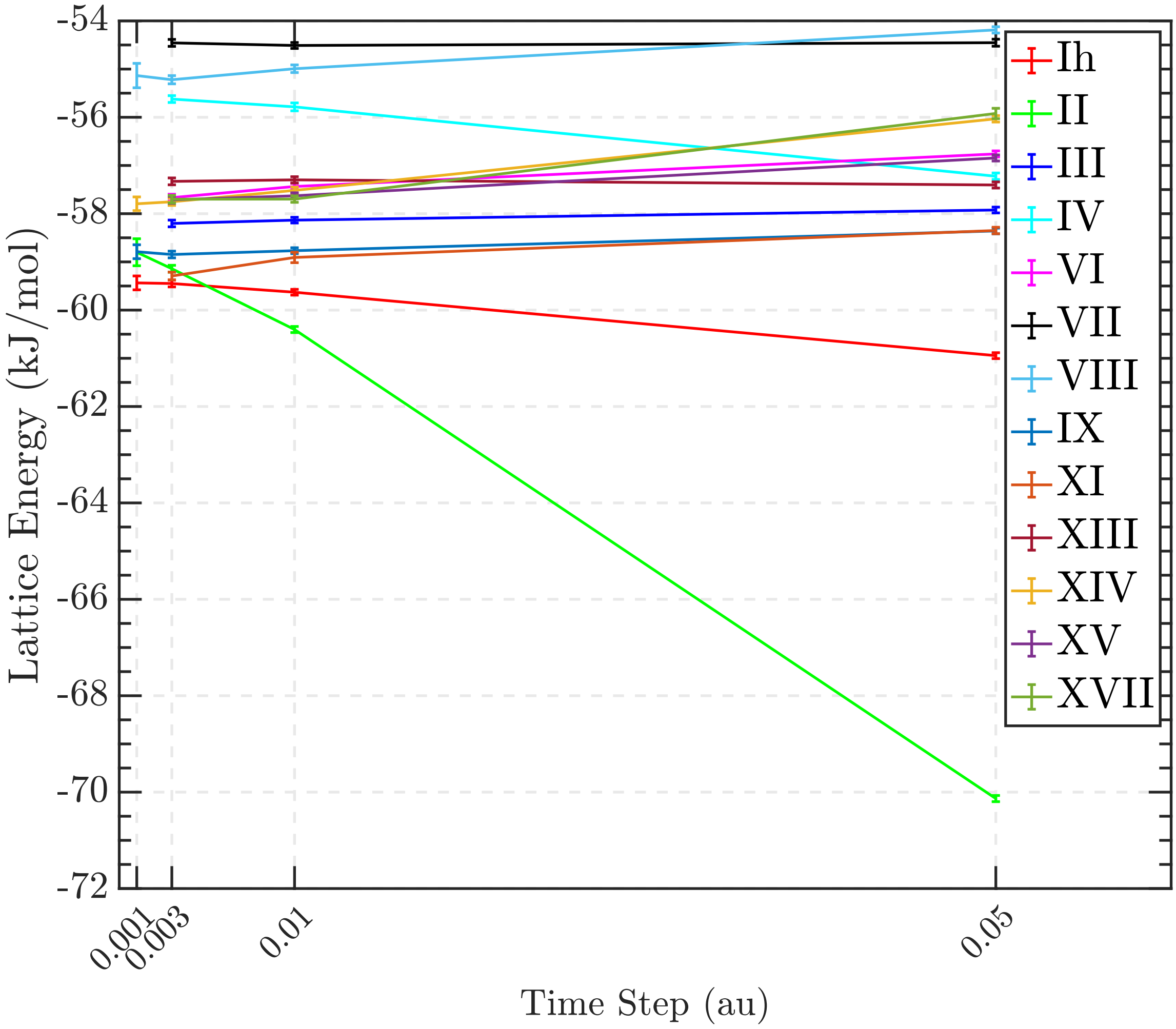

Figure 3: Time step convergence of the absolute lattice energy: all the curves are reported on a single energy scale to showcase the differences among all the polymorphs.

# 7 DFT benchmark in meV

In Table 8 we provide the exact same Table 2 of the main manuscript. Energies are now given in meV, so that the reported numbers are directly comparable to the VASP output.

| Method | Ih | II | III | IV | VI | VII | VIII | IX | XI | XIII | XIV | XV | XVII | MAE |
|---|---|---|---|---|---|---|---|---|---|---|---|---|---|---|
| DMC | -616 | -613 | -603 | -576 | -597 | -564 | -582 | -610 | -614 | -594 | -598 | -598 | -598 | |
| | | | | | | | | | | | | | | |
| B3LYP-D4 | -653 | -633 | -626 | -602 | -608 | -557 | -570 | -636 | -655 | -624 | -618 | -608 | -643 | 23 |
| B3LYP-D3(BJ)$^{atm}$ | -659 | -641 | -633 | -610 | -618 | -570 | -583 | -643 | -661 | -632 | -626 | -617 | -650 | 30 |
| B3LYP-D3(BJ) | -664 | -648 | -639 | -617 | -626 | -579 | -592 | -649 | -666 | -639 | -634 | -625 | -653 | 36 |
| B3LYP-D3$^{atm}$ | -665 | -664 | -644 | -631 | -645 | -621 | -632 | -657 | -666 | -653 | -649 | -644 | -655 | 51 |
| B3LYP-D3 | -669 | -670 | -649 | -638 | -653 | -630 | -641 | -663 | -671 | -660 | -656 | -652 | -659 | 58 |
| B3LYP | -546 | -487 | -500 | -454 | -444 | -373 | -387 | -503 | -546 | -474 | -461 | -446 | -540 | 123 |
| | | | | | | | | | | | | | | |
| revPBE0-D4 | -584 | -560 | -556 | -530 | -535 | -486 | -498 | -564 | -585 | -550 | -543 | -535 | -573 | 51 |
| revPBE0-D3(BJ)$^{atm}$ | -590 | -570 | -565 | -541 | -548 | -501 | -515 | -573 | -591 | -561 | -555 | -548 | -579 | 40 |
| revPBE0-D3(BJ) | -594 | -577 | -570 | -548 | -556 | -511 | -524 | -579 | -596 | -568 | -563 | -556 | -583 | 33 |
| revPBE0-D3$^{atm}$ | -591 | -590 | -568 | -558 | -574 | -567 | -578 | -581 | -592 | -577 | -574 | -574 | -582 | 21 |
| revPBE0-D3 | -595 | -597 | -574 | -565 | -582 | -576 | -587 | -587 | -596 | -584 | -581 | -582 | -585 | 17 |
| revPBE0 | -457 | -389 | -409 | -360 | -345 | -268 | -283 | -411 | -456 | -377 | -363 | -348 | -449 | 218 |
| | | | | | | | | | | | | | | |
| PBE0-MBD | -675 | -654 | -648 | -623 | -629 | -580 | -592 | -657 | -678 | -644 | -638 | -629 | -664 | 43 |
| PBE0-TS | -670 | -665 | -649 | -635 | -648 | -434 | -606 | -661 | -673 | -657 | -654 | -647 | -659 | 59 |
| PBE0-D4 | -667 | -637 | -637 | -606 | -609 | -553 | -565 | -644 | -670 | -627 | -619 | -608 | -656 | 29 |
| PBE0-D3(BJ)$^{atm}$ | -670 | -640 | -641 | -610 | -613 | -558 | -571 | -647 | -673 | -631 | -623 | -613 | -660 | 31 |
| PBE0-D3(BJ) | -675 | -647 | -646 | -617 | -621 | -568 | -581 | -653 | -677 | -638 | -631 | -620 | -663 | 37 |
| PBE0-D3$^{atm}$ | -682 | -671 | -657 | -634 | -642 | -600 | -611 | -667 | -685 | -655 | -649 | -641 | -672 | 55 |
| PBE0-D3 | -686 | -671 | -663 | -641 | -649 | -609 | -621 | -673 | -689 | -662 | -657 | -648 | -675 | 61 |
| PBE0 | -596 | -540 | -552 | -508 | -502 | -434 | -448 | -554 | -597 | -527 | -516 | -502 | -588 | 69 |
| | | | | | | | | | | | | | | |
| SCAN+rVV10 | -707 | -703 | -687 | -675 | -687 | -664 | -679 | -698 | -710 | -696 | -693 | -690 | -697 | 95 |
| R2SCAN | -651 | -636 | -626 | -607 | -617 | -584 | -597 | -632 | -654 | -626 | -622 | -616 | -644 | 27 |
| RSCAN | -636 | -615 | -608 | -586 | -594 | -558 | -570 | -613 | -639 | -605 | -600 | -593 | -630 | 10 |
| SCAN | -668 | -651 | -642 | -622 | -628 | -597 | -612 | -651 | -672 | -642 | -637 | -632 | -660 | 43 |
| | | | | | | | | | | | | | | |
| optB88-vdW | -704 | -703 | -693 | -680 | -694 | -658 | -670 | -703 | -708 | -701 | -700 | -693 | -688 | 97 |
| optB86b-vdW | -712 | -704 | -699 | -681 | -693 | -651 | -663 | -707 | -717 | -701 | -699 | -691 | -697 | 73 |
| optPBE-vdW | -679 | -681 | -671 | -661 | -672 | -635 | -647 | -683 | -682 | -681 | -679 | -672 | -662 | 95 |
| rev-vdW-DF2 | -688 | -666 | -666 | -641 | -648 | -600 | -612 | -672 | -693 | -661 | -656 | -646 | -676 | 59 |
| vdW-DF2 | -616 | -623 | -605 | -600 | -613 | -585 | -599 | -621 | -616 | -621 | -620 | -615 | -602 | 14 |
| vdW-DF | -555 | -557 | -548 | -539 | -547 | -509 | -523 | -562 | -555 | -558 | -556 | -549 | -537 | 51 |
| | | | | | | | | | | | | | | |
| revPBE-D4 | -621 | -585 | -592 | -560 | -559 | -504 | -515 | -594 | -625 | -577 | -570 | -559 | -611 | 26 |
| revPBE-D3(BJ)$^{atm}$ | -610 | -580 | -584 | -555 | -556 | -500 | -512 | -588 | -613 | -573 | -566 | -556 | -599 | 28 |
| revPBE-D3(BJ) | -614 | -586 | -590 | -562 | -564 | -509 | -521 | -594 | -618 | -580 | -574 | -564 | -602 | 22 |
| revPBE-D3$^{atm}$ | -607 | -592 | -582 | -563 | -574 | -559 | -568 | -589 | -610 | -581 | -576 | -573 | -598 | 14 |
| revPBE-D3 | -612 | -599 | -588 | -570 | -582 | -568 | -578 | -595 | -614 | -588 | -583 | -581 | -601 | 9 |
| revPBE | -455 | -369 | -404 | -344 | -320 | -227 | -241 | -399 | -456 | -359 | -342 | -323 | -447 | 236 |
| | | | | | | | | | | | | | | |
| PBE-MBD | -730 | -697 | -702 | -670 | -671 | -610 | -621 | -706 | -735 | -691 | -682 | -669 | -719 | 88 |
| PBE-TS | -721 | -702 | -699 | -677 | -683 | -610 | -626 | -705 | -726 | -698 | -692 | -681 | -710 | 90 |
| PBE-D4 | -722 | -679 | -690 | -653 | -650 | -583 | -594 | -692 | -726 | -672 | -663 | -649 | -711 | 71 |
| PBE-D3(BJ)$^{atm}$ | -725 | -684 | -694 | -657 | -655 | -589 | -601 | -695 | -730 | -676 | -667 | -653 | -715 | 76 |
| PBE-D3(BJ) | -729 | -690 | -699 | -664 | -663 | -598 | -610 | -702 | -734 | -683 | -675 | -661 | -718 | 82 |
| PBE-D3$^{atm}$ | -729 | -702 | -704 | -675 | -678 | -619 | -631 | -709 | -734 | -695 | -688 | -676 | -719 | 93 |
| PBE-D3 | -734 | -709 | -710 | -682 | -686 | -629 | -640 | -715 | -739 | -702 | -695 | -684 | -723 | 99 |
| PBE | -645 | -576 | -600 | -548 | -535 | -454 | -467 | -597 | -649 | -565 | -552 | -534 | -638 | 46 |
| | | | | | | | | | | | | | | |
| HF-D4 | -444 | -482 | -432 | -445 | -474 | -480 | -497 | -460 | -437 | -468 | -471 | -478 | -433 | 135 |
| HF-D3(BJ)$^{atm}$ | -524 | -555 | -512 | -519 | -550 | -557 | -572 | -535 | -518 | -542 | -547 | -552 | -513 | 59 |
| HF-D3(BJ) | -528 | -562 | -517 | -525 | -558 | -566 | -581 | -541 | -522 | -549 | -554 | -560 | -517 | 54 |
| HF-D3$^{atm}$ | -405 | -466 | -401 | -428 | -465 | -490 | -509 | -434 | -397 | -452 | -457 | -470 | -394 | 153 |
| HF-D3 | -409 | -472 | -406 | -435 | -473 | -500 | -518 | -440 | -401 | -459 | -465 | -477 | -397 | 146 |
| HF | -275 | -265 | -242 | -228 | -234 | -202 | -224 | -263 | -266 | -249 | -243 | -241 | -267 | 351 |
| | | | | | | | | | | | | | | |
| LDA | -1037 | -978 | -994 | -943 | -943 | -868 | -877 | -988 | -1049 | -964 | -953 | -936 | -1033 | 370 |

Table 8: Performance of each XC functional on the absolute lattice energy. Energies are reported in meV, so that they are directly comparable to the VASP output.

## 8 Water monomer in DMC-ICE13

In this section we provide details regarding the DMC and DFT simulation of the water monomer in DMC-ICE13. In detail, we report the time step behaviour of the DMC absolute energy in Table 9. Energies are reported in atomic units so that they are directly comparable to the CASINO output.

| $\tau$ | Energy |
|---|---|
| 0.001 | -17.21892(9) |
| 0.003 | -17.21884(2) |
| 0.01 | -17.21870(2) |
| 0.05 | -17.21937(2) |

Table 9: Time step dependence of the absolute energy for the water monomer in DMC-ICE13. Energies are in au, so that they are directly comparable to the CASINO output.

The single point DFT energies for all the considered functionals are reported in Table 10. Energies are given in eV so that they are directly comparable to the VASP output.

| Method | Energy |
|---|---|
| | |
| B3LYP-D4 | -17.3592 |
| B3LYP-D3(BJ)$^{atm}$ | -17.3664 |
| B3LYP-D3(BJ) | -17.3664 |
| B3LYP-D3$^{atm}$ | -17.3510 |
| B3LYP-D3 | -17.3510 |
| B3LYP | -17.3508 |
| | |
| revPBE0-D4 | -18.1958 |
| revPBE0-D3(BJ)$^{atm}$ | -18.2057 |
| revPBE0-D3(BJ) | -18.2057 |
| revPBE0-D3$^{atm}$ | -18.1825 |
| revPBE0-D3 | -18.1825 |
| revPBE0 | -18.1821 |
| | |
| PBE0-MBD | -18.1894 |
| PBE0-TS | -18.1836 |
| PBE0-D4 | -18.1878 |
| PBE0-D3(BJ)$^{atm}$ | -18.1911 |
| PBE0-D3(BJ) | -18.1911 |
| PBE0-D3$^{atm}$ | -18.1837 |
| PBE0-D3 | -18.1837 |
| PBE0 | -18.1836 |
| | |
| SCAN+rVV10 | -15.4477 |
| R2SCAN | -15.4055 |
| RSCAN | -16.1418 |
| SCAN | -15.6833 |
| | |
| optB88-vdW | -12.5518 |
| optB86b-vdW | -12.3949 |
| optPBE-vdW | -12.4780 |
| rev-vdW-DF2 | -12.7056 |
| vdW-DF2 | -12.9141 |
| vdW-DF | -12.5524 |
| | |
| revPBE-D4 | -14.2939 |
| revPBE-D3(BJ)$^{atm}$ | -14.2970 |
| revPBE-D3(BJ) | -14.2970 |
| revPBE-D3$^{atm}$ | -14.2688 |
| revPBE-D3 | -14.2688 |
| revPBE | -14.2681 |
| | |
| PBE-MBD | -14.2824 |
| PBE-TS | -14.2759 |
| PBE-D4 | -14.2811 |
| PBE-D3(BJ)$^{atm}$ | -14.2856 |
| PBE-D3(BJ) | -14.2856 |
| PBE-D3$^{atm}$ | -14.2759 |
| PBE-D3 | -14.2759 |
| PBE | -14.2758 |
| | |
| HF-D4 | -28.3882 |
| HF-D3(BJ)$^{atm}$ | -28.4843 |
| HF-D3(BJ) | -28.4843 |
| HF-D3$^{atm}$ | -28.3618 |
| HF-D3 | -28.3618 |
| HF | -28.3616 |
| | |
| LDA | -14.8850 |

Table 10: Single point DFT calculation for all the considered functionals on the water monomer. Energies are given in eV so that they are directly comparable to the VASP output.

# 9 Geometries of DMC-ICE13

In this section, we provide the geometries used in the DMC-ICE13 in the VASP POSCAR format.

Ice Ih
1.0
7.6780930000 0.0000000000 0.0000000000
3.8390460000 6.6494230000 0.0000000000
0.0000000000 0.0000000000 7.2345670000
H O
24 12
Direct
0.000007000 0.334718000 0.199799000
0.665252000 0.000010000 0.199780000
0.334714000 0.665261000 0.199791000
0.334760000 0.999976000 0.699786000
0.999980000 0.665239000 0.699800000
0.665240000 0.334755000 0.699799000
0.544461000 0.000006000 0.019584000
-0.000011000 0.455520000 0.019605000
0.455505000 0.544480000 0.019594000
0.455543000 0.999980000 0.519584000
0.999998000 0.544446000 0.519600000
0.544461000 0.455524000 0.519592000
0.332240000 0.879481000 0.984486000
0.211731000 1.120507000 0.984481000
0.879462000 0.788265000 0.984489000
0.788248000 0.332255000 0.984498000
0.667731000 0.211748000 0.984486000
0.120526000 0.211703000 0.484497000
0.667762000 1.120506000 0.484488000
0.788272000 0.879477000 0.484480000
0.211722000 0.667743000 0.484495000
0.332240000 0.788252000 0.484486000
-0.120503000 0.332222000 0.484499000
1.120486000 0.667750000 0.984491000
0.000009000 0.331330000 0.061616000
0.668637000 0.000012000 0.061595000
0.331326000 0.668660000 0.061608000
0.331370000 0.999972000 0.561601000
0.668637000 0.331348000 0.561616000
0.335676000 0.999992000 0.936746000
0.664294000 0.335703000 0.936768000
0.000012000 0.335655000 0.436771000
0.664333000 0.999995000 0.436740000
0.335670000 0.664304000 0.436760000
0.999979000 0.668634000 0.561617000
0.999974000 0.664306000 0.942299010

Ice II
1.0000000000000000
7.12333918 0.0 -3.04115415
-4.60506582 5.43464184 -3.04115415
0.0 0.0 7.74535894
H O
24 12
Cartesian
6.4179440000000003 10.2480430000000009 8.6935219999999980
6.6233660000000008 11.3038450000000026 9.8068299999999979
12.7683710000000001 7.4035430000000009 12.0511329999999983
13.0911720000000003 7.3198800000000013 13.5941349999999979
10.6468699999999998 6.1388150000000010 15.7181639999999980
11.9348799999999997 6.9781750000000011 15.8996309999999976
11.2707410000000010 9.1448300000000007 8.6282219999999974
12.5828120000000006 9.3371940000000002 7.7717469999999977
14.8027650000000008 6.7086010000000016 7.5624269999999978
14.8941400000000002 8.2338700000000014 7.3142819999999977
9.0242830000000005 9.1279330000000023 12.0845649999999978
7.9612280000000002 10.2533310000000011 11.7755999999999972
15.7107030000000005 7.0558020000000017 13.2061079999999968
15.5052810000000001 6.0000000000000009 12.0927999999999969
9.3602750000000015 9.9003030000000010 9.8484969999999983
9.0374750000000006 9.9839660000000023 8.3054949999999970
11.4817770000000010 11.1650300000000016 6.1814649999999975
10.1937660000000001 10.3256700000000023 5.9999999999999982
10.8579059999999998 8.1590160000000012 13.2714079999999974
9.5458350000000003 7.9666510000000015 14.1278819999999978
7.3258820000000000 10.5952440000000010 14.3372029999999988
7.2345070000000007 9.0699750000000012 14.5853479999999980
13.1043640000000003 8.1759120000000003 9.8150639999999978
14.1674190000000007 7.0505140000000015 10.1240299999999976
6.0000000000000000 10.5838740000000016 9.5328409999999977
12.3462240000000012 7.2895050000000010 12.9439559999999982
11.6254910000000002 6.0371280000000009 15.8726849999999970
12.1648710000000015 8.7490980000000018 8.4485499999999973
14.6660070000000005 7.3894510000000011 6.8485239999999976
8.5135230000000011 9.5541690000000017 11.3457849999999976
16.1286470000000008 6.7199710000000010 12.3667889999999971
9.7824230000000014 10.0143410000000017 8.9556739999999984
10.5031550000000014 11.2667170000000016 6.0269449999999978
9.9637760000000011 8.5547480000000018 13.4510789999999982
7.4626400000000004 9.9143940000000015 15.0511049999999980
13.6151240000000016 7.7496760000000009 10.5538449999999973

Ice III
1.00000000000000
6.6569609105993708 -0.0054159858429433 -0.0020629616267001
-0.0055656138125629 6.5590878313443408 -0.0421701683356358
0.0008280531233478 -0.0444073373963522 7.4520268810704300
H O
24 12
Direct
0.2416930182057721 0.3475829440608170 0.1816994472345053
0.7558983585359509 0.6580078849952085 0.6794428128909055
0.1436106184359576 0.7458705357531457 0.4249704596424581
0.2535285007030141 0.8437361046348452 0.0893958891186167
0.7393772822206327 0.1510649925514145 0.5854181504587973
0.6420017031025228 0.7549388056016562 0.3434532871313798
0.3667790895682086 0.9102800733489403 0.9126996753808975
0.6296347810697941 0.0782905647796893 0.4072150351139938
0.4216981124506174 0.1327859733809569 0.6598437425871636
0.5737617080457108 0.8559735723274845 0.1612553743988477
0.9905751980756208 0.6558058393515551 0.6727600731956899
0.1429253191628860 0.5081728172447544 0.4263324317971117
0.8596526768027072 0.5002974158599398 0.9222471742306149
0.1918715904968619 0.1188624002802925 0.7340854932346549
0.8884349114058958 0.1836982535564985 0.8466517241284141
0.3885670393874409 0.3142681836564602 0.9208228838301241
0.4007642071367766 0.5462563942883558 0.9965216219583005
0.5967780886789105 0.4400246432260620 0.4941061920371313
0.0969871048093069 0.0614940012398640 0.2687255585149823
0.4311445438383275 0.6094307532902067 0.5256244030287295
0.0068688429488458 0.3545818122531653 0.1762683074927703
0.6891631048616539 0.3841811192642142 0.0258548744973728
0.9318534681505469 0.8904815941252767 0.2380657197117454
0.9012302223339855 0.9530627753475723 0.7669944943880057
0.1209838136875883 0.3233422499913085 0.2585727943557895
0.8760231974108542 0.6863790516939207 0.7557836285880676
0.1704128115516534 0.6288254657897164 0.5061328879764992
0.8365746984248288 0.3807900429425403 0.0015192222260266
0.3724166040090466 0.8113686398846713 0.0126651198746325
0.6191107553406122 0.1778342990014059 0.5089557213613750
0.3362356084168784 0.1207882810904830 0.7689679713281051
0.6720569329811465 0.8704442469086680 0.2614726645959422
0.4276298588259164 0.4011973418998714 0.0262703192361813
0.5792272981624000 0.5896907552114885 0.5135415865576983
0.0792941646837134 0.9108362280618471 0.2538264328984859
0.9272347660774567 0.0993539431057401 0.7398368249979743

```
Ice IV
1.0
7.7817801603022598 0.1246969540912091 0.1590329264899646
2.7660103283375648 7.2746714077313941 0.1590329354147957
2.8239626713698587 1.9810340738602807 7.0171940989544641
H O
32 16
Direct
0.3267732012196201 0.8143038673813920 0.8869668263222142
0.4538777053407523 0.9681665658008156 0.7589371955710460
0.1664189933703033 0.1085393393011783 0.6792943660633546
0.5159200399071943 0.7211123901958476 0.6550490567214047
0.7596464436862322 0.4513350956205688 0.9794652888733071
0.6734562961138825 0.1707476473813579 0.1089913331666898
0.9890630136129168 0.2797250516927055 0.7378684890414060
0.6978944629345808 0.4457853308080580 0.8120044404274246
0.7797250280346449 0.4890630668689191 0.2378684890441700
0.6085393256658277 0.6664189327626067 0.1792943660534354
0.8579268950580138 0.7786466428688805 0.9767122459841655
0.7309455664065317 0.9899205188587846 0.2800099053285688
0.4681665093279234 0.9538778008180779 0.2589371955657597
0.4821497701334408 0.2868379348133108 0.3449816776440884
0.9807005372262345 0.7609848834138601 0.4480644889227515
0.1085974019852135 0.6774536424304048 0.1653619445120819
0.8140472956977037 0.6966787120536400 0.4414882447746983
0.1774537020360729 0.6085973565099668 0.6653619445457829
0.3143038256240746 0.8267732751975994 0.3869668263198978
0.0459314428258604 0.8094465320794648 0.6879675196725524
0.2609848868127839 0.4807005886890828 0.9480644889105159
0.3094465108188467 0.5459314692658449 0.1879675197248555
0.4899204661402866 0.2309456564103406 0.7800099053094054
0.6707476891419298 0.1734562221466749 0.6089913331565504
0.7868379769297223 0.9821497896985193 0.8449816775953451
0.1966786093296169 0.3140472925565648 0.9414882447776471
0.9513350274098802 0.2596464180265801 0.4794652888951136
0.2786466418670366 0.3579268893126860 0.4767122459551754
0.9457853572466640 0.1978944416956555 0.3120044404621204
0.2211123480642232 0.0159200203600997 0.1550490567686629
0.0549717823362939 0.0434182523125560 0.0696266113647722
0.5434181922720714 0.5549718423927089 0.5696266113503269
0.3720276298917963 0.8980566602373965 0.7548082158895950
0.7691494952448364 0.3634749078616249 0.9094299974921357
0.7389628330518332 0.5899409458462239 0.1256686759524074
0.6192723086047359 0.0898448582552103 0.2389050113961570
```

0.9093054613883381 0.7729361624311311 0.3602623181276357
0.1270334018119939 0.7395846096488602 0.5865772149713288
0.2395845156852393 0.6270334378599727 0.0865772149538210
0.5898448791026336 0.1192723803074721 0.7389050114244934
0.2729362340538460 0.4093055402744552 0.8602623181340661
0.0899410246466851 0.2389629047760348 0.6256686759439591
0.8634749439326040 0.2691494012957248 0.4094299974402092
0.3980566393899590 0.8720275581507618 0.2548082158965796
0.1074132695538457 0.1126692610348110 0.1096935934691168
0.3916342869768017 0.4039372654686496 0.3818654829436488
0.9039373140123111 0.8916342384104242 0.8818654829649225
0.6126692124051879 0.6074133181179779 0.6096935935293709

Ice VI
1.00000000000000
6.2270323974554032 -0.0163927873968318 0.0040454185171384
-0.0509592618291696 6.2278185213242514 0.0119537734634734
0.0341617017967277 -0.0197568446743032 5.8268993094045953
H O
20 10
Direct
0.4371253682351239 0.2433808704040077 0.0043181401638634
0.5220872052821197 0.7439848834595789 0.9827312352247526
0.0274103362134184 0.2384815935767541 0.0058056264599734
0.9318161576272260 0.7488359499070429 0.9812243001344662
0.2263486190987941 0.3607018885493740 0.3436432798740194
0.7328252217429474 0.6266491831157983 0.6434017856532785
0.2297370912658141 0.1139268863183508 0.3388631159547655
0.7294641649344615 0.8734288990577677 0.6481743223057298
0.2344038223726522 0.6699141641257942 0.6191094875842256
0.7248174597569740 0.3174283587633657 0.3679238136349658
0.6604563061858925 0.2463820505974746 0.8653909080964086
0.2987695882651243 0.7409562274185824 0.1216840946760095
0.3497664630237655 0.4523602364140202 0.6926486181617374
0.6094748616203074 0.5349977719139671 0.2943952337232009
0.3554776796063101 0.0366623377682139 0.6939900065367549
0.6037497254247717 0.9506774350298097 0.2930336250445976
0.1094460684556261 0.0268986565873535 0.6978020915288985
0.8497810641513897 0.9604526679049370 0.2892230143718022
0.0221490359800705 0.3561262775547866 0.7701404131449999
0.9370642250482468 0.6312026654923348 0.2168919093384802
0.2299421899652652 0.2386446884752011 0.2340013279547720
0.7292077607755308 0.7487104656283528 0.7530359278726776
0.2245381309586680 0.5109333461610743 0.6110556112434971
0.7346999447986390 0.4764081793234253 0.3759825134060344

0.2336447161950794 0.9602667454078624 0.6194586650291091
0.7255782650095581 0.0270907570421290 0.3675534034142349
0.9506341064672225 0.2383413563639687 0.8572501957482317
0.0085688938473227 0.7490010027193690 0.1297901632232094
0.5019450759668351 0.2459143944779663 0.8490212957142398
0.4572804517248374 0.7414200604413612 0.1380358747810569

Ice VII
1.00000000000000
10.5548693385330417 0.0000058980039804 0.0000003133324905
0.0000039320800553 7.0364774363851890 -0.0000001759153965
0.0000000174281243 -0.0000000357539034 3.3255328142960479
H O
24 12
Direct
0.0523313001160165 0.0784965280883925 0.1736101953456491
0.3856646334493450 0.0784965280883925 0.1736101953456491
0.7189979667826808 0.0784965280883925 0.1736101953456491
0.0523313001160165 0.5784965280883925 0.1736101953456491
0.3856646334493450 0.5784965280883925 0.1736101953456491
0.7189979667826808 0.5784965280883925 0.1736101953456491
0.2810020355016928 0.4215034918248377 0.1736101544559468
0.6143353688350285 0.4215034918248377 0.1736101544559468
0.9476687021683642 0.4215034918248377 0.1736101544559468
0.2810020355016928 0.9215034918248377 0.1736101544559468
0.6143353688350285 0.9215034918248377 0.1736101544559468
0.9476687021683642 0.9215034918248377 0.1736101544559468
0.2189974864148215 0.1715041982227929 0.6736117744894123
0.5523308197481572 0.1715041982227929 0.6736117744894123
0.8856641530814929 0.1715041982227929 0.6736117744894123
0.2189974864148215 0.6715041982227929 0.6736117744894123
0.5523308197481572 0.6715041982227929 0.6736117744894123
0.8856641530814929 0.6715041982227929 0.6736117744894123
0.1143358723014526 0.3284957576530858 0.6736116010402675
0.4476692056347883 0.3284957576530858 0.6736116010402675
0.7810025389681169 0.3284957576530858 0.6736116010402675
0.1143358723014526 0.8284957576530858 0.6736116010402675
0.4476692056347883 0.8284957576530858 0.6736116010402675
0.7810025389681169 0.8284957576530858 0.6736116010402675
0.9999999804683597 0.0000000265409881 0.9927788033850574
0.3333333138016954 0.0000000265409881 0.9927788033850574
0.6666666471350240 0.0000000265409881 0.9927788033850574
0.9999999804683597 0.5000000265409881 0.9927788033850574
0.3333333138016954 0.5000000265409881 0.9927788033850574
0.6666666471350240 0.5000000265409881 0.9927788033850574

0.1666666585309855 0.2499999976699172 0.4927774712836594
0.4999999918643141 0.2499999976699172 0.4927774712836594
0.8333333251976498 0.2499999976699172 0.4927774712836594
0.1666666585309855 0.7499999976699172 0.4927774712836594
0.4999999918643141 0.7499999976699172 0.4927774712836594
0.8333333251976498 0.7499999976699172 0.4927774712836594

Ice VIII
1.0000000000000000
4.84946537 0.0 0.0
0.0 4.84946537 0.0
0.0 0.0 7.05651379
H O
16 8
Cartesian
6.7815167500000006 7.5630335000000004 6.9666195000000002
9.2062492500000008 9.9877660000000006 10.4948765000000002
6.0000000000000009 9.2062492499999991 8.7307480000000002
9.9877660000000006 6.7815167499999998 12.2590050000000002
9.2062492500000008 7.5630335000000004 7.7641285000000000
9.2062492500000008 6.0000000000000000 7.7641285000000000
9.9877660000000006 9.2062492499999991 6.0000000000000000
6.0000000000000009 6.7815167499999998 9.5282570000000000
8.4247325000000011 9.2062492499999991 6.0000000000000000
8.4247325000000011 6.7815167499999998 12.2590050000000002
6.7815167500000006 6.0000000000000000 6.9666195000000002
7.5630335000000004 9.2062492499999991 8.7307480000000002
7.5630335000000004 6.7815167499999998 9.5282570000000000
9.2062492500000008 8.4247325000000011 10.4948765000000002
6.7815167500000006 9.9877660000000006 11.2923855000000000
6.7815167500000006 8.4247325000000011 11.2923855000000000
6.7815167500000006 6.7815167499999998 6.3680015000000001
6.7815167500000006 9.2062492499999991 8.1321300000000001
9.2062492500000008 6.7815167499999998 11.6603870000000001
9.2062492500000008 6.7815167499999998 8.3627465000000001
9.2062492500000008 9.2062492499999991 6.5986180000000001
6.7815167500000006 6.7815167499999998 10.1268750000000001
9.2062492500000008 9.2062492499999991 9.8962585000000001
6.7815167500000006 9.2062492499999991 11.8910035000000001

Ice IX
1.0

6.7849030000 0.0000000000 0.0000000000
0.0000000000 6.7849030000 0.0000000000
0.0000000000 0.0000000000 6.8079160000
H O
24 12
Direct
-0.0068300000 0.3400090000 0.1901610000
0.1139810000 0.1644120000 0.2833300000
0.3092700000 0.3703460000 0.1062840000
0.1599910000 0.4931700000 0.4401610000
0.3355880000 0.6139810000 0.5333300000
0.1296540000 0.8092700000 0.3562840000
0.8400090000 0.5068300000 -0.0598390000
0.6644120000 0.3860190000 0.0333300000
0.8703460000 0.1907300000 0.8562840000
0.5068300000 0.8400090000 1.0598390000
0.3860190000 0.6644120000 0.9666700000
0.1907300000 0.8703460000 0.1437160000
0.4931700000 0.1599910000 0.5598390000
0.6139810000 0.3355880000 0.4666700000
0.8092700000 0.1296540000 0.6437160000
0.3703460000 0.3092700000 -0.1062840000
0.6599910000 1.0068300000 0.3098390000
0.8355880000 0.8860190000 0.2166700000
0.6296540000 0.6907300000 0.3937160000
1.0068300000 0.6599910000 0.6901610000
0.8860190000 0.8355880000 0.7833300000
0.6907300000 0.6296540000 0.6062840000
0.3400090000 -0.0068300000 0.8098390000
0.1644120000 0.1139810000 0.7166700000
0.1140880000 0.3100300000 0.2673660000
0.1899700000 0.6140880000 0.5173660000
0.8100300000 0.3859120000 0.0173650000
0.3859120000 0.8100300000 0.9826340000
0.6140880000 0.1899700000 0.4826340000
0.6899700000 0.8859120000 0.2326340000
0.8859120000 0.6899700000 0.7673660000
0.3100300000 0.1140880000 0.7326340000
0.4042560000 0.4042560000 0.0000000000
0.0957440000 0.9042560000 0.2500000000
0.9042560000 0.0957440000 0.7500000000
0.5957440000 0.5957440000 0.5000000000

Ice XI
1.0000000000000000

4.4121733262067568 0.0000000000000000 0.0000000000000000
0.0000000000000000 7.6634880949752553 0.0001243032473494
0.0000000000000000 0.0001170212828762 7.1995921130067480
H O
16 8
Direct
0.0000000000000000 0.5407220845666505 0.0198885747150915
0.0000000000000000 0.6643169898827676 0.1999179185039158
0.0000000000000000 0.4592799382830144 0.5198865178764480
0.0000000000000000 0.3356761666602813 0.6999111051240845
0.4999990847117601 0.1643172156105884 0.1999208555514642
0.4999990847117601 0.0407255382696221 0.0198933812545808
0.4999990847117601 0.8356695688624748 0.6999237541050638
0.4999990847117601 0.9592740557354967 0.5198968235632565
0.3184513355265226 0.7681136336238013 0.9837204820447121
0.6815475203631769 0.7681136336238013 0.9837204820447121
0.3184568008982993 0.2318861246002663 0.4837188604109399
0.6815422838134533 0.2318861246002663 0.4837188604109399
0.1815451214512125 0.2681154376102950 0.9837171249536550
0.8184537344384810 0.2681154376102950 0.9837171249536550
0.1815444702737954 0.7318897939694392 0.4837161574284184
0.8184543856159046 0.7318897939694392 0.4837161574284184
0.0000000000000000 0.6656938065869755 0.0607742510025692
0.0000000000000000 0.3343051326678330 0.5607681524620171
0.4999990847117601 0.1656959204420655 0.0607760287173637
0.4999990847117601 0.8342987265895413 0.5607806058326680
0.4999990847117601 0.8328906902663383 0.9364367994468514
0.4999990847117601 0.1671212840622064 0.4364135329084324
0.9999990847117601 0.3328865700852581 0.9364216725851086
0.9999990847117601 0.6671210596289391 0.4364196457121191

Ice XIII
1.0
8.7223970000 0.0000000000 -3.1208890000
0.0000000000 7.4903640000 0.0000000000
0.0000000000 0.0000000000 10.3217530000
H O
56 28
Direct
0.3433870000 0.6439070000 0.2995050000
0.2477030000 0.4925390000 0.3343290000
0.1208210000 0.7159080000 0.1606500000
0.4707670000 0.9109100000 0.3499100000
0.5688800000 0.7497390000 0.4413980000
-0.0396450000 0.7528240000 0.0502080000

0.2317350000 0.4511680000 0.0951080000
0.3130960000 0.4901570000 0.5784070000
0.1782290000 0.3572010000 0.5167960000
0.2974000000 0.3643430000 -0.0146230000
0.0797290000 0.0142030000 0.1978620000
0.3184880000 0.1381530000 0.2492310000
0.3811580000 0.0489320000 0.1434190000
0.0461390000 0.2054080000 0.2507640000
0.1566130000 0.1439070000 0.7004950000
0.2522970000 -0.0074610000 0.6656710000
0.3791790000 0.2159080000 0.8393500000
0.0292330000 0.4109100000 0.6500900000
-0.0688800000 0.2497390000 0.5586020000
0.5396450000 0.2528240000 0.9497920000
0.2682650000 0.9511680000 0.9048920000
0.1869040000 0.9901570000 0.4215930000
0.3217710000 0.8572010000 0.4832040000
0.2026000000 0.8643430000 1.0146230000
0.4202710000 0.5142030000 0.8021380000
0.1815120000 0.6381530000 0.7507690000
0.1188420000 0.5489320000 0.8565810000
0.4538610000 0.7054080000 0.7492360000
0.6566130000 0.3560930000 0.7004950000
0.7522970000 0.5074610000 0.6656710000
0.8791790000 0.2840920000 0.8393500000
0.5292330000 0.0890900000 0.6500900000
0.4311200000 0.2502610000 0.5586020000
1.0396450000 0.2471760000 0.9497920000
0.7682650000 0.5488320000 0.9048920000
0.6869040000 0.5098430000 0.4215930000
0.8217720000 0.6427990000 0.4832040000
0.7026000000 0.6356570000 1.0146230000
0.9202710000 0.9857970000 0.8021380000
0.6815120000 0.8618480000 0.7507690000
0.6188420000 0.9510680000 0.8565810000
0.9538610000 0.7945920000 0.7492360000
0.8433870000 0.8560930000 0.2995050000
0.7477030000 1.0074610000 0.3343290000
0.6208210000 0.7840920000 0.1606500000
0.9707670000 0.5890900000 0.3499100000
1.0688800000 0.7502610000 0.4413980000
0.4603550000 0.7471760000 0.0502070000
0.7317350000 0.0488320000 0.0951080000
0.8130960000 0.0098430000 0.5784070000
0.6782280000 0.1427990000 0.5167960000
0.7974000000 0.1356570000 -0.0146230000
0.5797290000 0.4857970000 0.1978620000
0.8184880000 0.3618480000 0.2492310000

0.8811580000 0.4510680000 0.1434190000
0.5461390000 0.2945920000 0.2507640000
0.2578110000 0.5594060000 0.2548620000
0.4646290000 0.8020160000 0.4021910000
0.0609100000 0.8098980000 0.0969580000
0.2725670000 0.4014010000 0.5028720000
0.2018140000 0.4018270000 0.0005220000
0.4160680000 0.1075450000 0.2351890000
0.1244450000 0.1101530000 0.2656190000
0.2421890000 0.0594060000 0.7451380000
0.0353710000 0.3020160000 0.5978090000
0.4390900000 0.3098980000 0.9030420000
0.2274330000 0.9014000000 0.4971280000
0.2981860000 0.9018270000 0.9994780000
0.0839320000 0.6075450000 0.7648110000
0.3755550000 0.6101530000 0.7343810000
0.7421890000 0.4405940000 0.7451380000
0.5353710000 0.1979840000 0.5978090000
0.9390900000 0.1901020000 0.9030420000
0.7274330000 0.5985990000 0.4971280000
0.7981860000 0.5981730000 0.9994780000
0.5839320000 0.8924550000 0.7648110000
0.8755550000 0.8898470000 0.7343810000
0.7578110000 0.9405940000 0.2548620000
0.9646290000 0.6979840000 0.4021910000
0.5609100000 0.6901020000 0.0969580000
0.7725670000 0.0985990000 0.5028720000
0.7018140000 0.0981730000 0.0005220000
0.9160680000 0.3924550000 0.2351890000
0.6244450000 0.3898470000 0.2656190000

Ice XIV
1.0
8.3681120000 0.0000000000 0.0000000000
0.0000000000 8.1568520000 0.0000000000
0.0000000000 0.0000000000 4.0914040000
H O
24 12
Direct
0.5327190000 0.8367290000 0.4827050000
0.0917500000 0.2046200000 0.2616330000
0.7864760000 -0.0258540000 0.9070950000
0.7267500000 0.4682170000 0.3199870000
0.4115490000 0.5864610000 0.3636450000
0.8443400000 0.3269500000 0.3998320000

0.4082500000 0.7953800000 0.7616330000
0.0884510000 0.4135390000 0.8636450000
-0.0327190000 0.1632710000 -0.0172950000
0.7732500000 0.5317830000 0.8199870000
0.6556600000 0.6730500000 0.8998320000
1.0327190000 0.6632710000 0.5172950000
0.9115490000 -0.0864610000 0.6363550000
0.5884520000 1.0864610000 0.1363550000
0.9082500000 0.7046200000 0.2383670000
0.2732500000 0.9682170000 0.1800130000
0.1556600000 0.8269500000 0.1001680000
0.2864760000 0.5258540000 0.0929050000
0.7135240000 1.0258540000 0.4070950000
0.2267500000 0.0317830000 0.6800130000
0.3443400000 0.1730500000 0.6001680000
0.4672810000 0.3367290000 1.0172950000
0.2135240000 0.4741460000 0.5929050000
0.5917500000 0.2953800000 0.7383670000
0.0065520000 0.2531210000 0.1245720000
0.6402580000 0.9880400000 0.2320320000
0.2465370000 0.8884100000 0.0050460000
0.4934480000 0.7468790000 0.6245720000
0.2534630000 0.1115900000 0.5050460000
0.8597420000 0.0119600000 0.7320320000
0.7465370000 0.6115900000 0.9949540000
0.7534630000 0.3884100000 0.4949540000
0.9934480000 0.7531210000 0.3754280000
0.3597420000 0.4880400000 0.2679680000
0.1402580000 0.5119600000 0.7679680000
0.5065520000 0.2468790000 0.8754280000

Ice XV
1.0
6.2584630000 0.0000000000 0.0010920000
0.0087560000 6.2700020000 -0.0065670000
0.0000000000 0.0000000000 5.8146080000
H O
20 10
Direct
0.2469380000 0.3651950000 0.3511890000
0.2507480000 0.1200220000 0.3466180000
0.2544860000 0.6738570000 0.6259200000
0.6800220000 0.2512950000 0.8728700000
0.3688750000 0.4565470000 0.7003470000
0.3755430000 1.0399450000 0.7014680000

0.1308850000 1.0334030000 0.7050020000
1.0406450000 0.3614220000 0.7769500000
0.4588680000 0.2479420000 1.0117830000
1.0486020000 0.2443360000 1.0125810000
0.5411320000 0.7520580000 -0.0117830000
-0.0486020000 0.7556640000 -0.0125810000
0.7530620000 0.6348050000 0.6488110000
0.7492520000 0.8799780000 0.6533820000
0.7455140000 0.3261430000 0.3740800000
0.3199780000 0.7487050000 0.1271300000
0.6311250000 0.5434530000 0.2996540000
0.6244570000 -0.0399450000 0.2985320000
0.8691150000 -0.0334030000 0.2949980000
-0.0406450000 0.6385780000 0.2230500000
0.2541040000 0.9654530000 0.6272630000
0.9714750000 0.2444850000 0.8637490000
0.2511690000 0.2438870000 0.2416610000
0.2443130000 0.5163730000 0.6179820000
0.5224680000 0.2503160000 0.8563990000
0.7488310000 0.7561130000 0.7583390000
0.7556870000 0.4836270000 0.3820190000
0.4775320000 0.7496840000 0.1436010000
0.0285250000 0.7555150000 0.1362520000
0.7458960000 0.0345470000 0.3727380000

Ice XVII
1.00000000000000
3.1167530443944949 5.3761745952595694 -0.0204149081686747
-3.0998500347949216 5.3866022682842472 -0.0293056841941682
-0.0087214532495416 0.0276468375045884 5.9428093147058343
H O
12 6
Direct
0.9174182150804928 0.1318558269267445 0.1950481747149136
0.0036720979434577 0.9399164566265364 0.5283985023930514
0.1094406339310997 0.0262413296267588 0.8617403185566139
0.7129890622874933 0.0583421054668458 0.2606488190595944
0.2814174406082114 0.7353674327769991 0.5940235567081172
0.0361514583519138 0.3040245351406234 0.9273634976553200
0.6547090098016021 0.4256693960498537 0.4314566443957404
0.9722231916499646 0.6773392907884199 0.7647700068625337
0.4034131358347107 0.9946861265582808 0.0981690221140879
0.5730319835916708 0.7893306044366578 0.2802423814533805
0.6903021839055404 0.5953789159458464 0.6135598520589951
0.7670120470606219 0.7127292233406267 0.9468918825897781

0.1090076922341698 0.1319086937308709 0.9892341235594616
0.8121994595927635 0.1313119504912416 0.3225214399846704
0.1093262958763062 0.8347336316406275 0.6559110740233568
0.5780067254383119 0.5990632106467652 0.4909295497283178
0.8755525807923762 0.6005326415603316 0.8242228997491368
0.5768267860193259 0.8978786282459675 0.1576082543929021